\documentclass{aa}  

\usepackage{graphicx}
\usepackage{txfonts}
\usepackage{hyperref}
\hypersetup{
        colorlinks   = true,
        citecolor=blue,
}
\usepackage[flushleft]{threeparttable}

\begin{document}
        
        \title{Fine structure in the luminosity function in young stellar populations with \textit{Gaia} DR2}
        
        \author{Difeng Guo
                \inst{1}\thanks{difengguo.astro@gmail.com}
                \and
                Alex de Koter
                \inst{1,2}
                \and
                Lex Kaper
                \inst{1}
                \and
                Anthony G.A. Brown
                \inst{3}
                \and
                Jos H.J. de Bruijne
                \inst{4}
        }
        \institute{Anton Pannekoek Institute for Astronomy, University of Amsterdam, Science Park 904, 1098 XH, Amsterdam
                \and
                Instituut voor Sterrenkunde, Celestijnenlaan 200D bus 2401,
                3001 Leuven, Belgium
                \and
                Leiden Observatory, Leiden University, Niels Bohrweg 2, 2333 CA Leiden, The Netherlands
                \and 
                European Space Agency (ESA), European Space Research and Technology Centre (ESTEC), Keplerlaan 1, 2201 AZ Noordwijk, The Netherlands}
        
        \date{Received September 00, 0000; accepted March 00, 0000}
        \authorrunning{Guo et al.} 
        \titlerunning{The fine structure in the LF in \textit{Gaia} DR2}
        
        \abstract
        {
            A pioneering study showed that the fine structure in the luminosity function (LF) of young star clusters contains information about the evolutionary stage (age) and composition of the stellar population.
            The notable features include the H-peak, which is the result of the onset of hydrogen burning turning pre-main sequence stars into main sequence stars.
            The feature moves toward the faint end of the LF, and eventually disappears as the population evolves. Another detectable feature is the Wielen dip, a dip at $M_V$ $\simeq$ 7 mag in the LF first identified in 1974 for stars in the solar environment. Later studies also identified this feature in the LF of star clusters. The Wielen dip is caused by the increased importance of H$^-$ opacity in a certain range of low-mass stars.
        }
        {
            We studied the detailed structure in the luminosity function using the data from \textit{Gaia} DR2 and PARSEC stellar evolution models with the aim to further our understanding of young stellar populations.
        }
        {
            We analyzed the astrometric properties of stars in the solar neighborhood ($< 20$ pc) and in various relatively nearby ($<$ 400 pc) young ($<$ 50 Myr) open clusters and OB associations, and compare the features in the luminosity function with those generated by PARSEC models.
        } 
        {
            The Wielen dip is confirmed in the LF of all the populations, including the solar neighborhood, at $M_G\simeq7$ mag. The H-peak is present in the LF of the field stars in the solar neighborhood. It likely signals that the population is mixed with a significant number of stars younger than 100 Myr. The H-peak is found in the LF of young open clusters and OB associations, and its location varies with age. The PARSEC evolutionary models predict that the H-peak moves from $\sim$ -1 mag towards $\sim$ +6 mag in $M_G$ for populations with  ages increasing from $1$ to $\sim 70$ Myr. Our observations with {\it Gaia} DR2 confirm the evolution of the H-peak from $\sim$5 Myr up to $\sim$47 Myr. We provide a calibration function between $M_{\rm G}$ and age that works in the age domain between 1 and 30 Myr.
        }
        {
            The fine structure in the luminosity function in young stellar populations can be used to estimate their age.}
        
        
        \maketitle
        
        \section{Introduction}
        \label{chap:LF|sec:introduction}
        
        The luminosity function (LF) is one of the prime observables of a stellar population, be it an open cluster, an OB association, or a mixture of field stars. The LF is defined as the distribution of brightness among the members of the population. The profile of the LF contains coded information on the initial mass function (IMF), age, and star formation history \citep{Luyten1968, Bessell1993}.
        The power of the LF lies in its simplicity; it only requires information on brightness, other than spectroscopic data, to infer a variety of fundamental properties of a stellar population, and can thus be studied in distant stellar populations. To accurately extract information from the LF, it is essential that all its characteristics are well understood. Towards the bright end, the LF contains fine structure that potentially provides additional diagnostic power, notably the H- and R-peaks \citep{Piskunov1996} and the Wielen dip \citep{Wielen1974}. 
        
        \cite{Piskunov1996} discussed two time-dependent features in the LF of a coevally forming population of stars, in which the more massive stars have already reached the main sequence (MS) and the lower mass stars are still in the pre-main sequence (PMS) phase. It is the details of the relation between mass and luminosity and between radius and luminosity of such a population that introduces the H-peak (or H-maximum) and R-peak (or R-maximum) in the LF. The H-peak is located at the brighter end of the LF and is linked to the arrival of stars on the MS, associated with a change in the slope of the mass-luminosity relation (MLR) creating a local peak in the stellar number density at a specific luminosity. The H-peak moves to the fainter end as the population evolves, as the mass of the stars entering the MS decreases over time. It disappears after $\sim$\,100\,Myr when stars of 0.6-0.7\,$M_{\odot}$ enter the MS for which the MLR is different. The R-peak lies at the fainter end of the LF and is short lived ($\lesssim 5$\,Myr). \cite{Piskunov1996} link it to a radius inversion in the regime of 1-2\,$M_{\odot}$ PMS stars, with radius decreasing with increasing mass for forming stars that just passed the birthline.
        
        The time evolution of the H-feature can be used as an independent method to estimate the age of a population. \citet{Belikov1997} and \citet{Piskunov2001} calibrated the correlation between the H-peak position in $M_V$ and age using stellar evolutionary models \citep{Iben1965, Dantona1985, Palla1993} and observational data from various literature sources \citep[see the list in][]{BelikovEtPiskunov1997}.
        
        The Wielen dip is a depression found in the LF at $M_V \simeq 7$, first reported by \cite{Wielen1974}. He counted stars in the Gliese catalog \citep[see][]{Gliese2015}, making a luminosity function for the field stars within 20 pc from the Sun. He compared the result with the empirical luminosity function by \citet{Luyten1968}, and identified a slight depression around $M_V \simeq 7$. 
        Further observational evidence of the Wielen dip is presented in various works. \cite{Haywood1994} discussed the luminosity function for Galactic disk stars including the Wielen dip; \cite{Lee1995} discovered the Wielen dip for the first time in a star cluster, the Pleiades, a finding independently confirmed by \cite{Belikov1998}. \cite{Jeffries2001} and \cite{Naylor2002} pointed out that the Wielen dip is present in the luminosity function of NGC 2516 and NGC 2457, respectively. More recent observational evidence is mentioned by \cite{Jao2018}, whose main focus was on a MS feature around $M_G \sim 10.7$ mag, where objects within 100 pc in the \textit{Gaia} DR2 catalog were investigated. \cite{Olivares2019} report Ruprecht 147 as the oldest cluster so far to feature the Wielen dip, although it is presented in the mass function instead of the luminosity function. 
        
        \cite{Lee1988} and \cite{Lee1995} proposed that the origin of the Wielen dip lies in a bi-modal time-dependent initial mass function. As the dip is seen in a diverse range of physical environments, the bi-modality should be a fundamental property of clumping and fragmenting of star-forming gas. Clearly, the IMF represents the outcome of the star formation process, and in principle may contain fine structure reflecting the detailed physics involved \citep[e.g.,][]{Bate2012}.
        \cite{Kroupa1990} provide a simpler physical explanation; they note that for decreasing stellar mass the importance of H$^-$ as a source of opacity increases, ultimately becoming the primary source. This occurs at a stellar mass of about $\sim$0.7\,$M_{\odot}$, flattening the local MLR and causing an
        indentation at $M_V \simeq 7$ in the luminosity function \citep[see also][]{Kroupa2002}.

        In this work we present observational evidence of the H-peak and Wielen dip in the nearby ($< 20$\,pc) field star population and in a number of young open clusters and OB associations based on \textit{Gaia} DR2 \citep{GaiaCollaboration2016, GaiaCollaboration2018a}. We discuss the relation between the age of the population and the position of the H-peak in the LF (the peak-age relation) and how this relation can be applied as an alternative age diagnostic. With the help of the PAdova and TRieste Stellar Evolution Code (PARSEC) evolutionary grid \citep{Bressan2012, Chen2014, Tang2014, Chen2015}, we calibrate the peak-age relation and assess how much it has improved since the work of \cite{Belikov1997} and \cite{Piskunov2001}.
        
        This paper is organized as follows. In Section~\ref{chap:LF|sec:lumfunc-20pc} we inspect the luminosity function in the solar neighborhood (< 20 pc) using different data sets to assess levels of completeness and their impact on the identification of the H-peak and the Wielen dip. In Section~\ref{chap:LF|sec:lumfunc-cluster} we introduce the stellar populations studied in this work. In Section~\ref{chap:LF|sec:mlr-evolution} we show the evolution of the MLR and its connection to the H-peak in the luminosity function, which leads to Section~\ref{chap:LF|sec:hpeak-evolution} where we discuss the time evolution of the H-peak and introduce an empirical relation between the H-peak position and age. Other discussions can be found in Section~\ref{chap:LF|sec:discussion}, and we conclude the paper in Section~\ref{chap:LF|sec:conclusion}.
        
        \begin{figure*}
                \centering
                \includegraphics[width=0.9\linewidth]{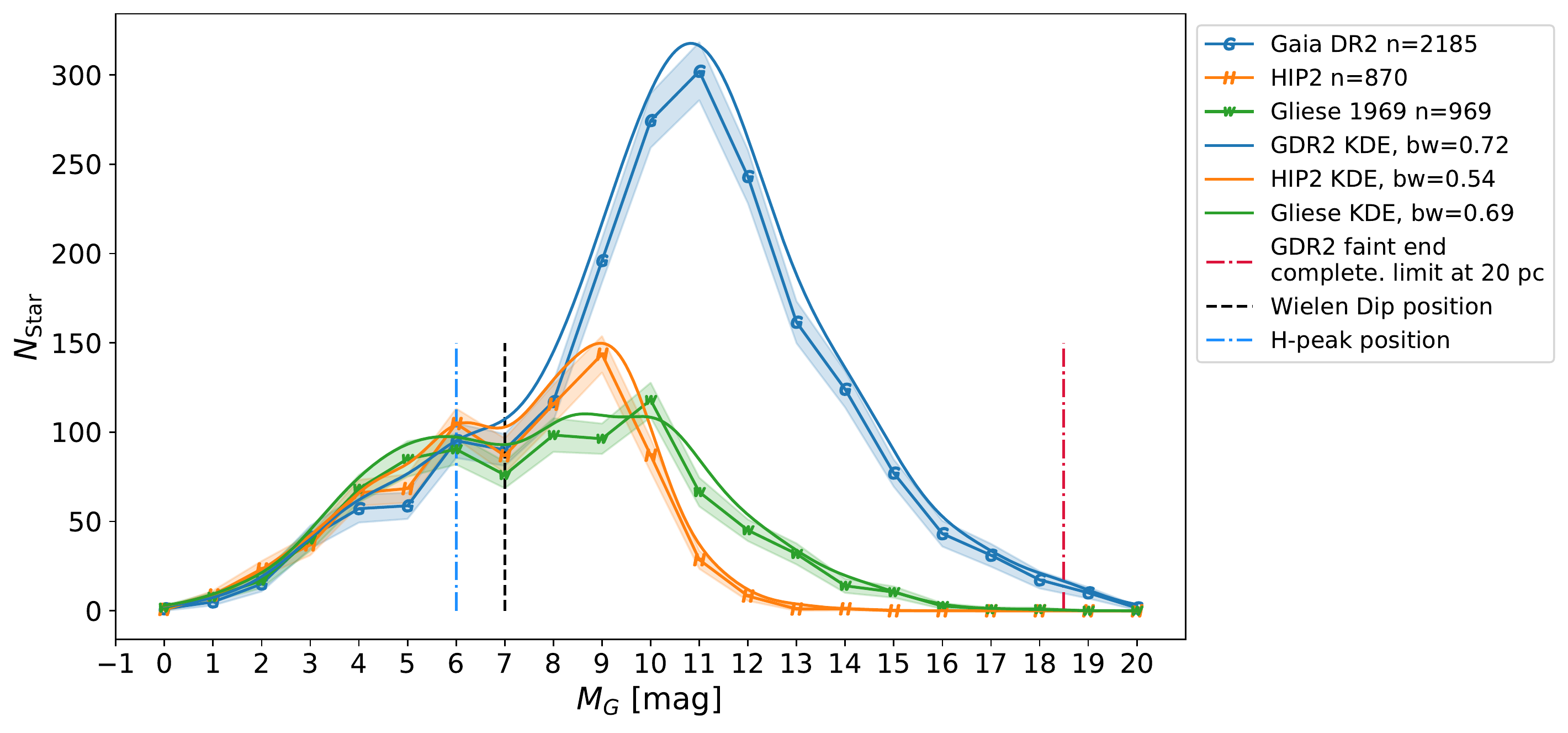}
                \caption{Luminosity function of stars within 20 pc obtained from three catalogs. The green curves are built from the  Gliese catalog, orange curves from the \textit{Hipparcos} HIP2 catalog, and blue curves from \textit{Gaia} DR2; the colored areas indicate the estimated uncertainty of the histogram based on a bootstrap method. The smooth curves were made using the kernel density estimation method. The location of the H-peak and Wielen dip are indicated. Binning and incompleteness issues are discussed in section~\ref{chap:LF|sec:lumfunc-20pc}. 
                }
                \label{fig:lumfunction-20pc}
        \end{figure*}

        \section{Luminosity function in the solar neighborhood}
        \label{chap:LF|sec:lumfunc-20pc}
        The stars in the solar vicinity represent a rich and diverse population dominated by, on average, low-mass stars of several gigayears in age 
        \citep[e.g.,][]{Binney2000, Hinkel2017}, but also interspersed, at least at distances up to $\sim$150\,pc, with dynamically distinct groups of young stars \citep[e.g.,][]{Gagne2018, Binks2020, Gagne2020}. Their proximity allows precise distance determinations and
        a detailed understanding of their astrophysical properties without great concern for extinction and reddening. 
        
        In this section we inspect the luminosity function of the population within 20\,pc in \textit{Gaia}\,DR2, and compare it with the Gliese \citep{Gliese1995} and {\it Hipparcos} \citep{VanLeeuwen2007} catalogs. We adopted the same parallax limit (parallax $\geq$ 50 mas) in \textit{Gaia} DR2 as that in \citet{Wielen1974} for easier comparison. The \texttt{ADQL} code to select the sample from \textit{Gaia} DR2 in the {\it Gaia} Archive is provided in the Appendix.
        
        Immediately after obtaining the sample, we applied the necessary photometric corrections following the recipes of \cite{Evans2018} and \cite{Apellaniz2018}. During this process, one bright source (source ID: Gaia DR2 6165699748415726848)\footnote{This source is identified as $\iota$~Centauri by SIMBAD.} showed an abnormal behavior due to incorrect flux in the BP and RP bands. As its color $(G - G_{\rm RP}) < -20$\,mag is far from the normal sources, we exclude this source from the results of this work and do not show it in the color-magnitude diagram.
        
        Because of the close distance, for the majority of the sources the ratio of the parallax to its error ($\varpi / \sigma_\varpi$) is above 5 in all three catalogs. Thus, for simplicity we did not apply any quality filter on parallax which may affect the completeness, and we were able to estimate the distance from inverting the parallax without significant bias \citep{Bailer-Jones2015} for sources in \textit{Gaia} DR2 and \textit{Hipparcos}. The sources from the Gliese catalog have on average lower $\varpi / \sigma_\varpi$, with some sources below 5, thus we apply the exponential prior function for distance inference to this catalog \citep[see Section 7 of ][]{Bailer-Jones2015}, and use  the inferred distance instead of inverted parallax in this work.
                
        In Fig.~\ref{fig:lumfunction-20pc} we show the luminosity function of stars within 20 pc from the three catalogs applying the same bin size of 1 mag as in \citet{Wielen1974}. The green curve  reproduces the LF in \cite{Wielen1974}, who relied on the 1969 edition of Gliese catalog \citep[see][]{Gliese1969, Gliese1991, Gliese2015}. To produce this curve we limit ourselves to Gliese's objects in the 1969 version of the catalog and converted the V-band magnitude with the provided (V - I) color index to the G-band magnitude in \textit{Gaia}\,DR2 according to the recipe in \citet{Evans2018} so that we can construct a LF in $M_{G}$. In the original work of \cite{Wielen1974}, the stars in Gliese's catalog fainter than $\mathrm{M_V} \sim$ 7.5 mag do not reach completeness at the full 20\,pc distance, and the numbers in these bins were extrapolated from star counts at closer distances. In this work we do not perform the extrapolation, but construct the LF using the objects ``as is.'' The LF of Gliese's objects peaks at $M_G \simeq 10$ mag.
        
        The orange curve is made with all the sources within 20 pc in the Second Reduction of the \textit{Hipparcos} catalog (HIP2) \citep{VanLeeuwen2007}. This catalog achieves completeness within 20 pc for $\mathrm{M_V} < 10$\,mag \citep{Perryman1997}. As with Gliese's catalog, we converted the $H_p$ magnitude and (V - I) color index of the \textit{Hipparcos} catalog to the G magnitude in \textit{Gaia}\,DR2 in order to construct a LF in $M_G$, which peaks at $M_G \simeq 9$ mag. 
        
        The blue curve is constructed from \textit{Gaia}\,DR2, where the peak of the LF within 20\,pc is around $\mathrm{M_G} \simeq 11$ mag. \textit{Gaia} has completeness limits on both the faint (G $\sim$ 21) and bright (G $\sim$ 7) end \citep{Evans2018, Bouret2015}. The faint end limit poses no completeness issue for our study as the features of our interest are all at $M_G < 8$, which are far brighter than the limit ($M_G \sim 18.5$ at 20 pc) at such close distance. It is the bright end limit (the limit in $M_G$ ranges from 8.5 to 5.5 for a distance from 5 to 20 pc) that affects our observation, as within 20 pc some sources would be too bright to be recorded properly by $Gaia$. Therefore, despite its superior accuracy compared to the Gliese and HIP2 catalogs, \textit{Gaia} DR2 does not achieve completeness at the bright end of the LF ($M_G < 8$); thus, the H-peak and the Wielen dip may not be properly presented with the \textit{Gaia} catalog. We attempt to compensate for the incompleteness by filling the {\it Gaia} catalog with HIP2 sources that do not have a counterpart in the \textit{Gaia} catalog using the cross-match table by \citet{Marrese2018}. This added 91 sources to the sample. The combined sample makes up the blue curve in Fig.~\ref{fig:lumfunction-20pc}.
        
        At the bright end ($M_G < 8$), all three histograms show the Wielen dip at $M_G = 7$, as in \cite{Wielen1974}, as well as a peak next to it at $M_G = 6$. In the $M_G = 5$ bin the curves of HIP2 and \textit{Gaia} DR2 show a dip, while the Gliese curve does not; the bins at $M_G$ = 3 and 4 also contain more stars in the  Gliese sample than in the other two samples. The \textit{Gaia} DR2 sample, even after having compensated for the incompleteness at the bright end with HIP2 sources, has the fewest   stars in the bins $M_G$ = 4, 5, and 6. This discrepancy indicates how incompleteness can affect our interpretation of the luminosity function.
        In addition, we also recognize that the choice of the bin size (1 mag) is arbitrary, and is only there to reproduce the result of \cite{Wielen1974}. To this end, we also apply kernel density estimation  (KDE; see section~\ref{chap:LF|subsec:LF_clusters} for details) to properly construct the luminosity function from the three samples. The KDE curves are plotted as smooth curves in the same color as the respective histograms. In the KDE curves the dip of the HIP2 and \textit{Gaia} samples at $M_G = 5$ is smoothed out; the peak at $M_G \simeq 6$ and the Wielen dip remain present in the curves of Gliese and HIP2, while all the features are smoothed out in the \textit{Gaia} sample.
        
        The peak at $\mathrm{M_G} \simeq 6$ in the Gliese and HIP2 samples can be identified with the H-peak. This peak is expected to remain visible in stellar populations up to an age of 100~Myr (see Sect.~\ref{chap:LF|sec:introduction}), signifying that the solar neighborhood indeed contains an appreciable young stellar population. There is no indication of the R-peak, which is only expected for populations young enough to show significant ongoing star formation.
        
        At the faint end ($M_G > 8$) the \textit{Gaia} sample produces a high peak at $M_G = 11$, dominating the other two samples fading out due to completeness limits of the surveys at the faint end. The large discrepancy (100 -- 200 sources per bin) between \textit{Gaia} and the other two samples at the faint end demonstrates \textit{Gaia}'s great sensitivity.
        
\begin{figure}
                \centering
                \includegraphics[width=1.0\linewidth]{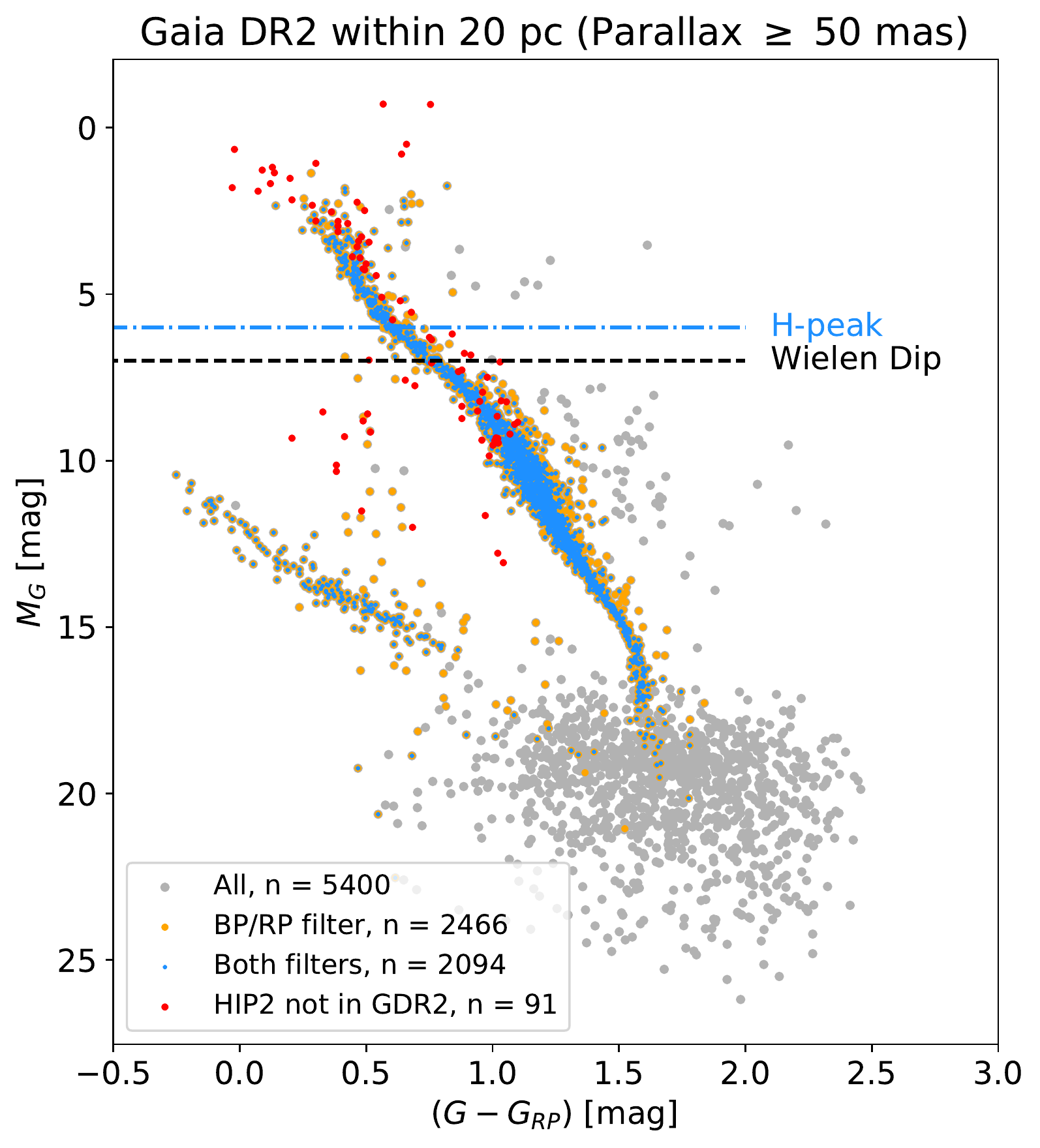}
                \caption{{\it Gaia} DR2 objects within 20 pc  plotted in the color--absolute magnitude diagram. The gray points are all objects within 20 pc; the orange points are all objects that pass the BP/RP filter; the blue points  are objects that pass both the BP/RP excess filter and the RUWE $<$ 1.4 filter (the sample used to make the blue curve in Fig.~\ref{fig:lumfunction-20pc}).
                The gray points form a clump below $M_G \simeq 15$ mag. These have unreliable astrometry and photometry. On the blue side of the main sequence the white dwarf cooling sequence is apparent.} 
                \label{fig:cmdsolarneighborhood20pcfilters}
        \end{figure}

        The color-magnitude diagram of the combined \textit{Gaia} sample is shown in Fig.~\ref{fig:cmdsolarneighborhood20pcfilters}; the raw sample from the {\it Gaia} Archive contains 5400 sources (gray points). The number is inflated to twice the estimated $\sim$2500 stars within 20 pc \citep[see, e.g.,][]{Stauffer2010} because of spurious astrometric and photometric solutions in \textit{Gaia} DR2. Although the parallax of the raw sample has been accurately measured ($\varpi / \sigma_{\varpi}$ well above 10), about half of the stars are excluded from our {\it Gaia} sample for the luminosity due to the poor quality of the photometry \citep[BP/RP excess filter][see Appendix~\ref{chap:LF|appendix:data_select_gaia_dr2} for more details]{Evans2018}. The majority of these objects with unreliable photometry are fainter than $M_G = 15$ mag, which corresponds to the brown dwarf regime \citep[][]{Pecaut2013}\footnote{\url{http://www.pas.rochester.edu/~emamajek/EEM\_dwarf\_UBVIJHK\_colors\_Teff.txt}}. There are 2446 stars that pass the BP/RP filter, which is consistent with the $\sim2500$ estimation for the solar neighborhood. They are colored orange in the plot. In addition to the BP/RP excess filter, we further exclude stars with unreliable overall astrometry data using the {re-normalized unit weight error} (RUWE) following the description in \citet{LL:LL-124} \citep[see also][]{Lindegren2018}, which mainly excludes sources of $M_G > 15$. The stars that pass both filters are colored blue. Because the two filters significantly reduce the number of stars, our final sample of 2094 stars in this work is only half-way from including all the 5400 sources within 20~pc distance. Most of the excluded sources are at the faint end. Lastly, the additional sources from HIP2 without a \textit{Gaia} counterpart are colored red. These sources are mainly distributed at $M_G < 10$, showing the effect of \textit{Gaia}'s bright end completeness limit at close distance. Together with the sources colored blue, they make up the \textit{Gaia} DR2 sample shown in Fig.~\ref{fig:lumfunction-20pc}.
        
        \section{The luminosity function in young open clusters and OB associations}
        \label{chap:LF|sec:lumfunc-cluster}
        
        Although the H-peak feature is present in the solar neighborhood field stars, it is better observed in coeval young stellar populations, as demonstrated in the work of \cite{Piskunov2004}, or older populations with a limited age spread such as the Pleiades \citep{Belikov1998}. In this work we take advantage of the high accuracy of the \textit{Gaia} astrometric and photometric information of nearby stellar populations to perform a comparative study of the LF fine structure, notably the H-peak.
        
        \subsection{Selection of samples}
        
        We established the detailed membership, age, and extinction of the young open clusters and the Scorpius-Centaurus OB2 (Sco~OB2) association included in the Scorpius-Canis Majoris (ScoCMa) stream \citep{Bouy2015}, the nearest region of (massive) star formation (Guo et al., in prep. and Guo et al., in prep.) 
        This sample served to study the LF of young stellar populations spanning an age range of 10 -- 50~Myr. Added to this sample is the young open cluster NGC 6231 in Scorpius OB1 at $\sim 1.6$ kpc (van der Meij, in press A\&A 2021). 
        
        There are several reasons for choosing these populations. Most of these moving groups are no farther than 350~pc from the Sun, which means that we have a better chance of obtaining a well-resolved luminosity function. Furthermore, we have studied them in detail so that we already have the information of membership, age (by fitting PARSEC
        isochrones), and LF at hand, and we know they are all younger than 50 Myr with slight to moderate interstellar extinction ($A_V < 1.1$ mag) \citep[see, e.g.,][]{Pecaut2016}. We only include the cluster in our sample if the LF resolution is sufficient (requiring more than 100 stars) and the H-peak is detectable by eye. Furthermore, as an attempt to include more young open clusters between 20 and 30 Myr, we considered the membership results of \citet{Cantat-Gaudin2018b} and the age determination of \citet{Bossini2019}. We were able to find a few clusters in this age range; however, the achievable magnitude resolution of the LF in these clusters is not suited for this work due to the small number of stars, the large distance, or high extinction. NGC 6231 is a special case for very young clusters ($\sim 5$ Myr). Although it is much more distant ($\sim$1.6 kpc) than the other clusters and consequently has a limited magnitude range of the LF, it still provides a rare example showing the position of the H-peak in the LF of a very young population. Last but not least, these clusters are at such a distance that both \textit{Gaia}'s bright and faint end completeness limits do not affect the analysis.
        
        \subsection{The luminosity function of 14 young stellar populations}
        \label{chap:LF|subsec:LF_clusters}
        
\begin{figure*}
                \centering
                \includegraphics[width=0.9\linewidth]{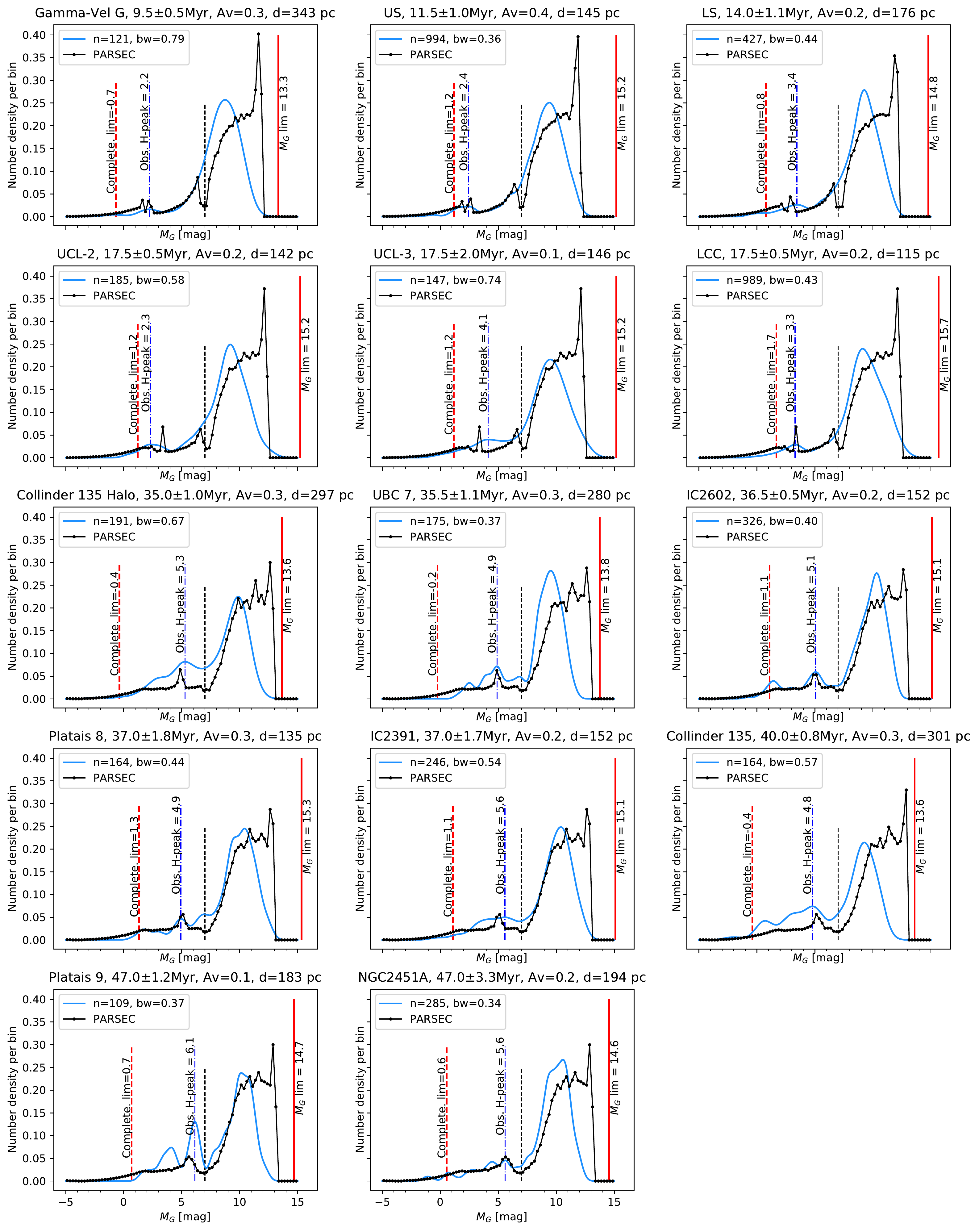}
                \caption{Luminosity function for stars with a membership probability greater than 0.95 of young open clusters and Sco~OB2 in the ScoCMa stream, ordered by increasing age.
                        The black line is the theoretical luminosity function from PARSEC, at the same age as the cluster's isochrone age. The bin size is 0.25 mag. A vertical scaling factor of 3 is applied to fit with the observational profiles. The blue line is the observed LF made by KDE. The kernel bandwidths are labeled  ``bw'' 
                        in the insets. The PARSEC LFs show the H-peak at $M_G$ $2 \sim 6$ mag and the Wielen dip at $\sim 7 $ mag; the observed LFs show the H-peak, while the Wielen dip is sometimes smoothed out by the KDE procedure (compare the LF of LCC and Collinder 135). For reference, the black vertical dashed line indicates $M_G$ = 7 mag, the position of the Wielen dip. The blue vertical dot-dashed line gives the H-peak position in the observed LF. The red solid line indicates the faint end observational limit of \textit{Gaia} DR2 at the cluster's distance, and the red dashed line represents the completeness limit at the bright end of \textit{Gaia} DR2. All LF features of interest in this work are within the completeness limit.
                }
                \label{fig:lumfunctionclusters}
        \end{figure*}

        The luminosity function of the selected young open clusters in the ScoCMa stream and association subgroups in Sco~OB2 are presented in Figure~\ref{fig:lumfunctionclusters}, together with PARSEC LFs of the same isochronal age (see Section~\ref{chap:LF|sec:mlr-evolution} for details regarding these models). The PARSEC models used in this work follow the parameters chosen in Table~\ref{tab:PARSEC_input_parameters}, including an IMF of \cite{Chabrier2001} (for mass $< 1 M_\odot$) and \cite{Salpeter1955} (for mass $\geq 1 M_\odot$). The selected populations are, in sequence of ascending age, $\gamma$-Vel G \citep[named by][]{Cantat-Gaudin2019a}, Upper Scorpius (US), Lower Scorpius (LS), Upper Centaurus Lupus-2 (UCL-2, a subclump in UCL), UCL-3 (the same naming pattern as UCL-2), Lower Centaurus Crux (LCC), Collinder 135 Halo (the less dense population surrounding Collinder 135 and UBC 7), UBC 7, IC 2602, Platais 8, IC 2391, Collinder 135, Platais 9, and NGC 2451A. The cluster ages are the result of isochrone fitting with the method used in \cite{Joergensen2005} and \cite{Cantat-Gaudin2019b}. Our method fits the isochrone for both age and extinction ($A_V$). The extinction information provided by \textit{Gaia} DR2 is not complete, and it is not trivial to convert between $A_G$ and $A_V$ \citep[see][]{Evans2018, Andrae2018}. In order to estimate the uncertainty of the age, we carry out a bootstrapping process. For each population, we randomly redraw 90\% of its members (by assuming 10\% to be interlopers) and obtain the age-extinction determination repetitively (200 times). We take the median of the collection of results as the adopted age and extinction, and the standard deviation of the collection as the estimated uncertainty due to interloper contamination. Furthermore, as our isochrone grid step size is 0.5 Myr for age and 0.1 mag for $A_V$, any estimated uncertainties smaller than the grid step size were replaced by the step size as adopted uncertainties.
        
        Relevant information on the sample clusters is listed in Table~\ref{tab:cluster_info}. In the table the second column lists the number of stars identified as members in Guo et al. (in prep.) and Guo et al. (in prep.), 
        and the third column lists the number of stars that have reliable photometry and are used in the isochrone fitting process; 
        see the table caption for further information.
        
        \begin{table*}          
                \centering
                \caption{Open clusters and OB associations studied in this paper and whose  LFs are plotted in Figs.~\ref{fig:lumfunctionclusters} and \ref{fig:lumfunctionngc6231}.
                }
                \label{tab:cluster_info}
                \begin{threeparttable}
                \begin{tabular}{lrrrrrrrr}
                        \hline\\[-8pt]
                        Name &  $N_{\rm tot}$ &$N_{\rm phot}$ &  $\varpi$ &  $\sigma_{\varpi}$ &   Age [Myr] &  $\sigma_{\rm Age}$ &   $A_V$ &  $\sigma_{A_V}$ \\[2pt]
                        \hline\\[-8pt]
                        $\gamma$-Vel G &   134 &121 &  2.9 &      0.1 &   9.5 &    0.5 &  0.3 &   0.1 \\
                        US &  1137 &994 &  6.9 &      0.5 &  11.5 &    1.0 &  0.4 &   0.1 \\
                        LS &   487 &427 &  5.8 &      0.2 &  14.0 &    1.1 &  0.2 &   0.1 \\
                        UCL-2 &   202 &185 &  7.0 &      0.3 &  17.5 &    0.5 &  0.2 &   0.1 \\
                        UCL-3 &   157 &147 &  6.8 &      0.3 &  17.5 &    2.0 &  0.1 &   0.1 \\
                        LCC &  1048 &989 &  8.6 &      0.9 &  17.5 &    0.5 &  0.2 &   0.1 \\
                        Col 135 Halo &243    &191 &  3.4 &      0.3 &  35.0 &    1.0 &  0.3 &   0.1 \\
                        UBC 7 &   206 &175 &  3.6 &      0.1 &  35.5 &    1.1 &  0.3 &   0.1 \\
                        IC 2602 &   395 &326 &  6.6 &      0.3 &  36.5 &    0.5 &  0.2 &   0.1 \\
                        Platais 8 &   185 &164 &  7.4 &      0.3 &  37.0 &    1.8 &  0.3 &   0.1 \\
                        IC 2391 &   296 &246 &  6.6 &      0.5 &  37.0 &    1.7 &  0.2 &   0.1 \\
                        Col 135 &   200 &164 &  3.3 &      0.1 &  40.0 &    0.8 &  0.3 &   0.1 \\
                        Platais 9 &   111 &109 &  5.5 &      0.5 &  47.0 &    1.2 &  0.1 &   0.1 \\
                        NGC 2451A &   334 &285 &  5.1 &      0.2 &  47.0 &    3.3 &  0.2 &   0.1 \\
                        \hline\\[-8pt]
                        NGC 6231  &   268 & 267 &  0.61 &       0.04            &  4.7    &       0.4  &  1.7 &   0.1 \\
                        \hline& 
                
                \end{tabular}
                \begin{tablenotes}
                        \item Note: The meaning of the columns are, from left to right: cluster name, total number of members, number of members used in age determination, median parallax of the cluster in mas, standard deviation of the parallax in mas, the median isochronal age of the cluster in Myr, the uncertainty of the age in Myr, the estimated extinction $A_V$ in mag, and the uncertainty of $A_V$ in mag.
                \end{tablenotes}
                \end{threeparttable}
        \end{table*}

        A luminosity function is based on counting the number of stars within magnitude bins. In a plain histogram, the features in the LF can be affected by the size of the bins. Features can be smoothed out with a bin size that is too large, or overwhelmed by noise with a bin size that is  too small. In order to properly capture the features of the observed LF without an arbitrary choice of bin size, we build the LF profile using the KDE method\footnote{This work applies the method in the Python package \texttt{statsmodels}. See the documentation at \url{https://www.statsmodels.org/devel/generated/statsmodels.nonparametric.kernel_density.KDEMultivariate.html}} 
        which can automatically determine the bandwidth of the kernel function based on a given sample. The kernel function here is Gaussian. We select the highest peak at $M_G \leq 6.5$ mag as the position of the H-peak and take the kernel bandwidth as uncertainty. To locate the H-peak position predicted by the PARSEC model, we simply use the output luminosity function from the CMD tools website\footnote{\label{chap:LF|note:CMDtoolURL}\url{http://stev.oapd.inaf.it/cgi-bin/cmd}} (see the next section for more information) with a fixed bin size (0.25 mag) and identify the position of the highest peak at the high-brightness end as the H-peak, as the model is free from measurement errors of observations. The position of the H-peak in both observations and model predictions as a function of age is listed in Table~\ref{tab:hpeak_results}.

        \begin{table*}
                \centering
                \caption{H-peak position in the LF of the studied young clusters,  in observations and in model predictions.}
                \label{tab:hpeak_results}
                \begin{threeparttable}
                
                \begin{tabular}{lrrrrr}
                        \hline\\[-8pt]
                        Name &  Age [Myr] &  $\sigma_\mathrm{Age}$ &  H-peak &  BW &  H-peak model \\
                        \hline\\[-8pt]
                        $\gamma$-Vel G &      9.5 &        0.5 &        2.2 &      0.79 &             2.125 \\
                        US &     11.5 &        1.0 &        2.4 &      0.36 &             2.625 \\
                        LS &     14.0 &        1.1 &        3.4 &      0.44 &             2.875 \\
                        UCL-2 &     17.5 &        0.5 &        2.3 &      0.58 &             3.375 \\
                        UCL-3 &     17.5 &        2.0 &        4.1 &      0.74 &             3.375 \\
                        LCC &     17.5 &        0.5 &        3.3 &      0.43 &             3.375 \\
                        Col 135 Halo &     35.0 &        1.0 &        5.3 &      0.67 &             7.125 \\
                        UBC 7 &     35.5 &        1.1 &        4.9 &      0.37 &             6.375 \\
                        IC 2602 &     36.5 &        0.5 &        5.1 &      0.40 &             7.125 \\
                        Platais 8 &     37.0 &        1.8 &        4.9 &      0.44 &             6.375 \\
                        IC 2391 &     37.0 &        1.7 &        5.6 &      0.54 &             6.375 \\
                        Col 135 &     40.0 &        0.8 &        4.8 &      0.57 &             6.375 \\
                        Platais 9 &     47.0 &        1.2 &        6.1 &      0.37 &             5.625 \\
                        NGC 2451A &     47.0 &        3.3 &        5.6 &      0.34 &             5.625 \\
                        \hline \\[-8pt]
                        NGC 6231 &      4.7 &        0.4 &        1.1 &      0.38 &             0.875 \\
                        \hline
                \end{tabular}
                \begin{tablenotes}
                        \item Note: The first three columns (Name, Age, and $\sigma_{\rm Age}$) are identical to the corresponding columns in Table~\ref{tab:cluster_info}. Column 4 lists the position of the H-peak determined by the KDE method in $M_G$ magnitude, and Column 5 the bandwidth of the kernel function as uncertainty of the H-peak position. Column 6 is the position of the H-peak in the PARSEC model LF in $M_G$ magnitude. The model LF has a fixed bin size of 0.25 mag, and we use it as the uncertainty on the H-peak in the model.
                \end{tablenotes}
                
                \end{threeparttable}
        \end{table*}
        
        All the clusters have a luminosity profile (Fig.~\ref{fig:lumfunctionclusters}, blue lines) in close resemblance to the PARSEC profiles (black) of the corresponding isochrone at $2 < M_G < 9$ mag. The PARSEC LFs show that the position of the H-peak moves from 2 mag to 6 mag in $M_G$ as the population age increases from $\sim$10 Myr to $\sim$50 Myr. The major peaks of the luminosity function appear to indicate the completeness limit of the population in comparison to the {\it Gaia} completeness limit \citep{Evans2018, Boubert2020} at the cluster's distance (red vertical line). However, we cannot conclude that we have reached completeness for these clusters. 
        
        The PARSEC models cut off at $0.1 M_\odot$.
        The models drop sharply from the peak after 13 -- 14 mag due to the hard cut in the IMF, while the observed profiles peak at $\sim$10 mag and drop less dramatically. 
        This discrepancy is mostly due to the choice of the IMF for the model and the completeness of the population. The difference between the model and our observation at the faint end of the LF is not the focus of this work; the features of our interest lie in the brighter and well-populated section of the LF.
        
        See Fig.~\ref{fig:cmdclusters} in the Appendix for the positions of the Wielen dip and the H-peak in the color-magnitude diagram of the populations presented in this section.
        
        \subsection{The luminosity function of NGC 6231}
        \label{chap:LF|subsec:ngc6231}
        
\begin{figure}
                \centering
                \includegraphics[width=1.0\linewidth]{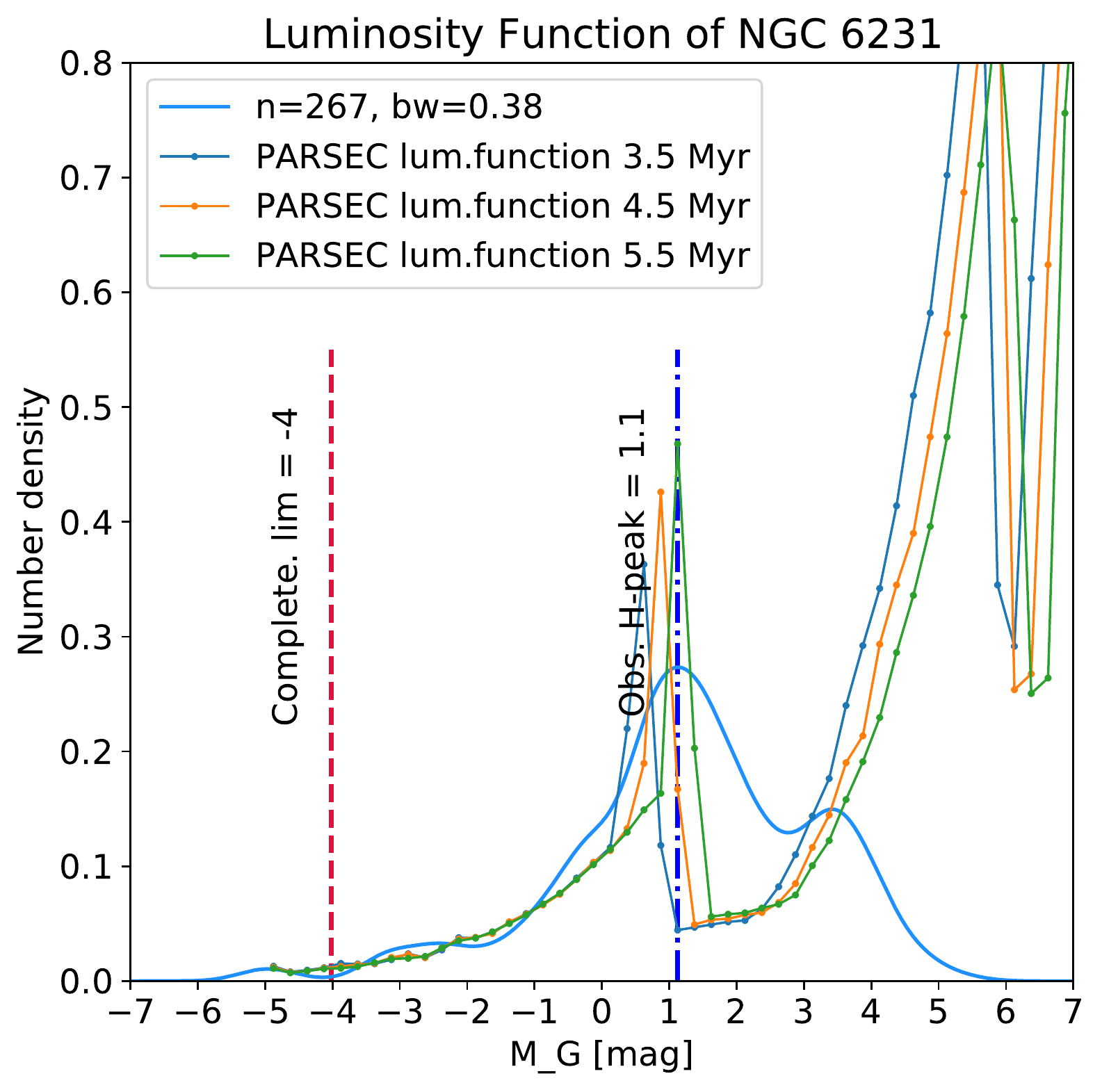}
                \caption{Luminosity function of NGC 6231 (blue curve) based on the KDE. The kernel bandwidth is 0.30. The thin blue, orange, and green histograms represent the PARSEC LF models for 3.5, 4.5, and 5.5~Myr in $M_G$. The models are vertically re-scaled by a factor of 30 to compare with the luminosity function of NGC 6231. The dash-dotted blue vertical line indicates the position of the observed H-peak at $M_G = 0.8$ mag; the dashed red vertical line indicates the completeness limit at the bright end of \textit{Gaia} DR2 at the distance of the cluster. The H-peak is within the completeness range of \textit{Gaia} DR2. The position of the H-peak is consistent with the age determination of NGC~6231: $4.7 \pm 0.4$~Myr (van der Meij, in press A\&A 2021).}
                \label{fig:lumfunctionngc6231}
        \end{figure}

        We included NGC 6231 in our analysis (van der Meij, in press A\&A 2021), a very young open cluster ($4.7\pm0.4$ Myr)  at a distance of $\sim1.6$~kpc. This is the youngest cluster that we have included in our study with sufficient data to resolve the bright end of the LF. The LF of the cluster is plotted in Fig.~\ref{fig:lumfunctionngc6231}. The faint end completeness limit of \textit{Gaia} beyond 1 kpc dictates that we can only reach completeness within $M_G < 5$ mag, with an extinction $A_V \simeq 1.5$ mag \citep{Sung2013}. The dip next to the H-feature shows up at $M_G \sim$ 4 mag for NGC 6231 before the extinction correction; if we correct for the extinction at an assumed $A_G \sim A_V \simeq 1.5$ mag as shown in the figure, then it is at $M_G \simeq$ 2.5, as predicted by the PARSEC luminosity function for 3.5 -- 5.5 Myr. The H-feature is located before the dip at $M_G\simeq1$ mag.
        
        \section{The evolution of the mass-luminosity relation in the PARSEC model}
        \label{chap:LF|sec:mlr-evolution}
        In this section we discuss how (transient) features in the luminosity function arise as a result of temporal changes in the mass-luminosity relation,
        which itself is a result of PMS stellar evolution.
        
        
        We use evolutionary tracks that cover the PMS phase from PARSEC models (version 1.2; for details see  \citealt{Chen2015}). Isochrones were obtained from the accompanying web page (see footnote \ref{chap:LF|note:CMDtoolURL}). The minimum initial mass covered by these tracks is $0.1 M_\odot$. The initial mass function (IMF) for low-mass stars is adopted from \citet{Chabrier2001}; for $M/M_\odot > 1$ we adopt \citet{Salpeter1955}. All relevant model parameters are summarized in Table~\ref{tab:PARSEC_input_parameters}.
        
        \begin{table*}[h!]
                \centering
                \caption{Input parameters for the PARSEC model isochrones and corresponding luminosity functions. }
                \label{tab:PARSEC_input_parameters}
                \begin{threeparttable}
                \begin{tabular}{r|l|c}
                        Option & Input & Default? \\ 
                        \hline\\[-18pt]
                        &&\\
                        PARSEC Version & 1.2S & Yes \\ 
                        
                        COLIBRI Version & S\_35 & Yes \\ 
                        
                        $n_{\rm inTPC}$ & 10 & Yes \\ 
                        
                        $\eta_{\rm Reimers}$ & 0.2 & Yes \\ 
                        
                        Photometric System & \textit{Gaia} DR2, cf. \citet{Apellaniz2018} & No \\ 
                        
                        Bolometric Corrections & YBC & Yes \\ 
                        
                        Circumstellar dust & All left as default & Yes \\ 
                        
                        Extinction & $\mathrm{A_V} $ = 0 mag & Yes \\ 
                        
                        Initial Mass Function & \citet{Chabrier2001} for $0.1 \leq M/M_\odot \leq 1$ & No \\ 
                        & \citet{Salpeter1955} for $M/M_\odot > 1$  & \\
                        
                        Ages & Linear age grid 1 Myr -- 70 Myr, step 0.5 Myr & No \\ 
                        
                        Metallicities & Z = 0.0152; fixed.       & Yes \\ 
                        
                        Output 1 & Luminosity function, range -5 -- 15 mag; bins 0.5 mag wide & No \\ 
                        
                        Output 2 & Isochrones & Yes \\ 
                        \hline 
                \end{tabular}
                        \begin{tablenotes}
                        \item Note: Parameters related to post-main sequence evolution are left as default values as they do not impact the results of this work. The parameter defaults only apply to version 3.3 of the CMD web tools; any updates of the web page may change the default options. $n_{\rm inTPC}$ is the resolution of the thermal pulse cycles in the COLIBRI section \citep[see][]{Marigo2017}; $\eta_{\rm Reimers}$ is the parameter of mass loss on the RGB using the Reimers formula; YBC refers to the bolometric correction database by \cite{Chen2019}. 
                        \end{tablenotes}
                        \end{threeparttable}
                 
        \end{table*}
        
        \cite{Piskunov1996} identified the H-feature to be associated with the non-monotonic 
        nature of the derivative of the MLR caused by PMS stars initiating core hydrogen burning and transitioning to the main sequence (MS) phase. Formally, the mass function
        \begin{eqnarray}
                f(m) = \frac{dN}{d\log m}
        \end{eqnarray}
        and the luminosity function
        \begin{eqnarray}
                \Phi(L) = \frac{dN}{d\log L} 
        \end{eqnarray}
        must conserve the total number of stars $N_{\rm tot}$,
        \begin{equation}
                \int dN = \int f(m)~d\log m = \int \Phi(L)~d\log(L) \equiv N_{\rm tot},
        \end{equation}
        where $dN$ is the number of stars in the logarithmic mass bin or the associated logarithmic luminosity bin being considered.  
        Consequently, the mass and luminosity function can be linked as
        \begin{equation}
                \Phi(L) = f(m) \left|\dfrac{d\log m}{ d\log L}\right|
        ,\end{equation}
        where $| \,d \log m / d\log L \,| = (L/m) \left| \, dm/dL \,\right|$ scales with the absolute value of the derivative of the MLR $m(L)$. It describes how many stars in a certain interval of mass are distributed to a certain interval of luminosity. As the absolute magnitude $M_G$ is related to luminosity via the bolometric correction, and the functions must stay non-negative, using the procedure similar to formula (2) in \cite{Belikov1997}, we can obtain the $M_G$ version of the relation
        \begin{equation}\label{eq:MLR-div}
                \Phi(M_G) = \frac{dN}{dM_G} = f(m) \left|\dfrac{d\log m}{d M_G}\right|.
        \end{equation}
        We  casually refer to this function as the luminosity function as well, and, in the same spirit, to $m(M_G)$ as the MLR.
        
        The observations and the theoretical models have both shown that $\Phi(M_G)$ is not monotonic, even if $f(m)$ is. It contains fine structure, implying that the derivative $\left| \,d\log m / d M_G\,\right|$   jumps up and down near local maxima or minima. Similarly, kinks in the mass function $f(m)$ also cause such jumps.
        
\begin{figure*}
                \centering
                \includegraphics[width=1.0\linewidth]{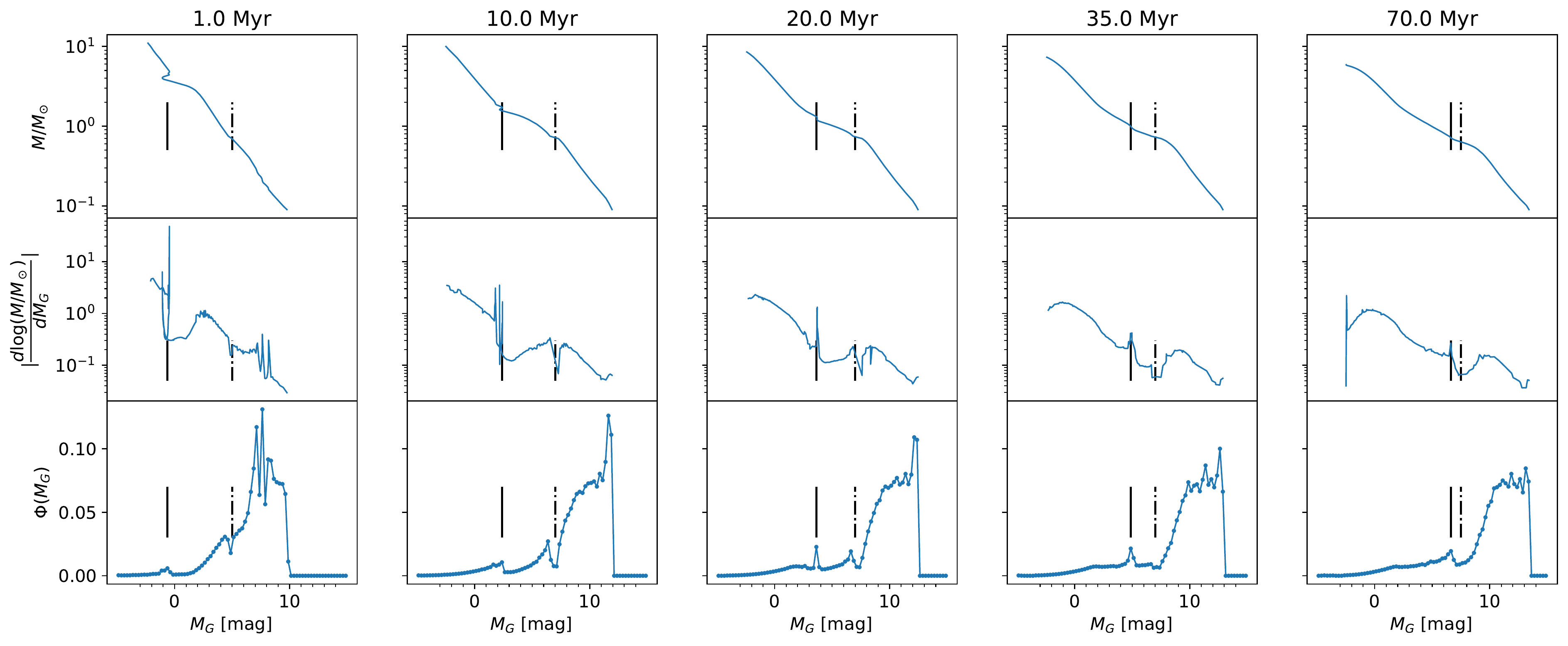}
                \caption{
                MLR, derivative of MLR, and LF from the PARSEC models in five snapshots. Top panels: MLR from the PARSEC model, as mass vs. $M_G$. Middle panels: Absolute value of the derivative $|\, d\log m / d M_G \,|$. Bottom panels: Luminosity functions from the PARSEC model. Each column is taken from a snapshot at a different time (indicated at the top) of the simulated population. In each panel the position of the H-peak is indicated by a vertical solid line, and the location of the Wielen dip by a vertical dot-dashed line.}
                \label{fig:mlr-div-lfrelation}
        \end{figure*}

        If we know how the shape of the MLR $m(M_G)$ evolves with time, then we can deduce the evolution of the luminosity function $\Phi(M_G)$ and features in this function using the relation in Eq.~\ref{eq:MLR-div}, for a known mass function $f(m)$. In Fig.~\ref{fig:mlr-div-lfrelation} we show $m(M_G)$ (top panel), the derivative of the logarithm of the mass with respect to $M_G$ (middle panel), and the luminosity function $\Phi(M_G)$ (lower panel) for PARSEC models at snapshots 1, 10, 20, 35, and 70 Myr. 
        
        Stars of about $4\,M_\odot$ evolve toward the main sequence in about 1\,Myr. At that age, the MLR shows a non-monotonic  section in the region $-1 < M_G < 0$ mag and $3 < M/M_\odot < 5$. The feature is reflected in the derivative curve as two sharp jumps (at $M_G \simeq -0.5$ and $-1.0$), where the derivative is approaching infinity, after which it loops back to a smoother behavior. It is  carried over to the luminosity function $\Phi(M_G)$ as a bump located at $-1 < M_G < 0$ and a flat section at $0 < M_G < 2$. The former is the H-peak. 
        
        Though the feature at $3 < M/M_\odot <5$ is the most prominent in the $m(M_G)$ relation at 1\,Myr, more subtle features, causing significant bumps and dips in the derivative are revealed as well. The subtle feature that is most prominently reflected in the luminosity function is a very narrow section of mass-luminosity flattening at $M_G \simeq 5$ (or $M/M_\odot \simeq 0.7$). This causes an identifiable dip in $\Phi(M_G)$. We return to the nature of this feature below.
        
        Using the PARSEC models we have learned so far that for an isochronal luminosity function at 1 Myr, save for the main peak at $6 < M_G < 10$ induced by the IMF and its cutoff at the low-mass end at 0.1\,$M_\odot$, we can identify at least three fine structures: the H-peak at $-1 < M_G < 0$, the `H-plateau' at $0 < M_G < 2$, and a dip at $M_G \simeq 5$ that we have not yet discussed.
        
        We can then try to observe the same features in the 10 Myr snapshot making the implicit assumption that the MLR changes continuously with time. We find that they are shifted toward the fainter end as the population evolves to activate core-hydrogen burning for ever lower-mass stars: the H-peak is now at $M_G \simeq 2$ mag, the H-plateau has narrowed to a small section near $M_G \simeq 3$ mag, and the as-yet-unidentified dip has moved to $M_G \simeq 7$ mag. Checking the panels for later evolutionary phases to assess any further evolution of the latter dip, we see that it no longer progresses (significantly) through the luminosity function, but remains stable at $M_G \simeq 7$ mag. We therefore identify it as the Wielen dip, found at this absolute magnitude in older populations. So, the Wielen dip forms first at $M_G \simeq 5$\,mag in very young clusters, moving to and stabilizing at $M_G \simeq 7$\,mag as late as 10\,Myr. 
        
        In the 10\,Myr snapshot the H-plateau has almost disappeared. In the 1\,Myr snapshot it is essentially the result from a flat section in the derivative. After this flat section, $|d\log M/dM_G|$ rises to reach a maximum at $M_G \simeq 2$\,mag, after which a downward slope starts all the way to the Wielen dip feature. This negatively sloped section has disappeared in the 10\,Myr snapshot, making the Wielen dip more prominent.
        Moving on to the 20 Myr snapshot, we find the H-peak at $M_G \simeq 4$\,mag and the Wielen dip unchanged at $M_G \simeq 7$\,mag, but somewhat less pronounced.
        
        The landscape of the LF starts to change in the snapshot of 35 Myr, as the ``loop back'' section of the mass-luminosity relation flattens and the magnitude difference between the location of the H-peak and the Wielen dip narrows. 
        At this stage, the movement of the H-peak stops being an effective age indicator, before its disappearance after 100-200 Myr. However, it may still be a signal of population mixture or age spread in older populations such as the Pleiades \citep{Belikov1998}. In the final snapshot of 70\,Myr, the H-peak ($M_G \simeq 6.5$ mag) is located closely to the Wielen dip ($M_G \simeq 7.5$ mag), making it indistinguishable from the boundary of the Wielen dip. The H-peak will evolve slowly together with the Wielen dip toward the faint end of the luminosity function, until their final disappearance once stars of $\sim 0.3\,M_{\odot}$ (at $\sim 160$ Myr) enter the end of pre-main sequence evolution, according to the PARSEC models.
        
        \section{The evolution of the H-feature}
        \label{chap:LF|sec:hpeak-evolution}
        
        In this section we use the location of the H-peak in the luminosity function to establish a chronometer of young clusters, as did \cite{BelikovEtPiskunov1997}.
        
        In Fig.~\ref{fig:hloc-agerelation} we summarize the relation between the H-peak's location in the luminosity function and the age of the population in the observed data (blue points) and in the  PARSEC models (black bars), as listed in Table~\ref{tab:hpeak_results}.
        In addition, we incorporate the age and H-peak information of six young open clusters (NGC 2383, NGC 2384, NGC 4103, Hogg 15, NGC 4755, NGC 7510) from \cite{Piskunov2004} (green points). The error bars for the H-peak location in $M_G$ are the uncertainty of converting the magnitude value from $M_V$ using the relation in \cite{Evans2018}, assuming that the color $(G_{BP} - G_{RP}) < 0.2$ as most of the stars are bright. 
        The horizontal dashed line at $M_G = 6$ mag indicates the predicted location of the H-peak in snapshots $> 40$ Myr, which coincides with the H-peak found in the solar environment (see Fig.~\ref{fig:lumfunction-20pc}), acting as an indicator of the asymptotic behavior of the H-peak.
        
        The overall impression of Fig.~\ref{fig:hloc-agerelation} is that there indeed exists a well-defined relation between the location of the predicted H-peak and age. The ages of the observed clusters in \textit{Gaia} DR2 result from isochrone fitting; they are in good agreement with those derived from the H-peak location, though at ages $>$30\,Myr the correspondence is somewhat less good. In conclusion, the H-peak is a good indicator of the age of young star-forming regions.
        
        \subsection{The evolutionary track critical points in the $M_G$--Age space}
        
        In addition to the various correlations we have established between the observational and model LFs on the evolution of the H-peak, the PARSEC evolutionary tracks can provide a deeper understanding of the physical origin of the feature.
        
        The PARSEC evolutionary tracks of \cite{Bressan2012} \footnote{See \url{https://people.sissa.it/~sbressan/parsec.html}. We do not use the Horizontal Branch models in this work} contain information on the evolutionary phase per model point, defined as critical points. For example, the first critical point is the  ``Beginning of PMS'' and is always the first point (at $t$ = 0 Myr) of any track. The phases related to the transition from PMS to MS are defined by three critical points, in sequence of time: ``End of PMS'', ``Near the zero age main sequence (ZAMS)'', and ``Beginning of MS.'' 
        The End of PMS phase is identified as the moment where rapid internal structure changes occur, causing a downturn in the evolutionary track \citep[or even loops; see][]{Bressan2012}. The ZAMS is defined as the point where the (blueward) evolutionary speed in the Hertzsprung-Russel diagram (HRD) 
        shows an abrupt drop by more than two orders of magnitude.
        
\begin{figure*}
                \centering
                \includegraphics[width=1.0\linewidth]{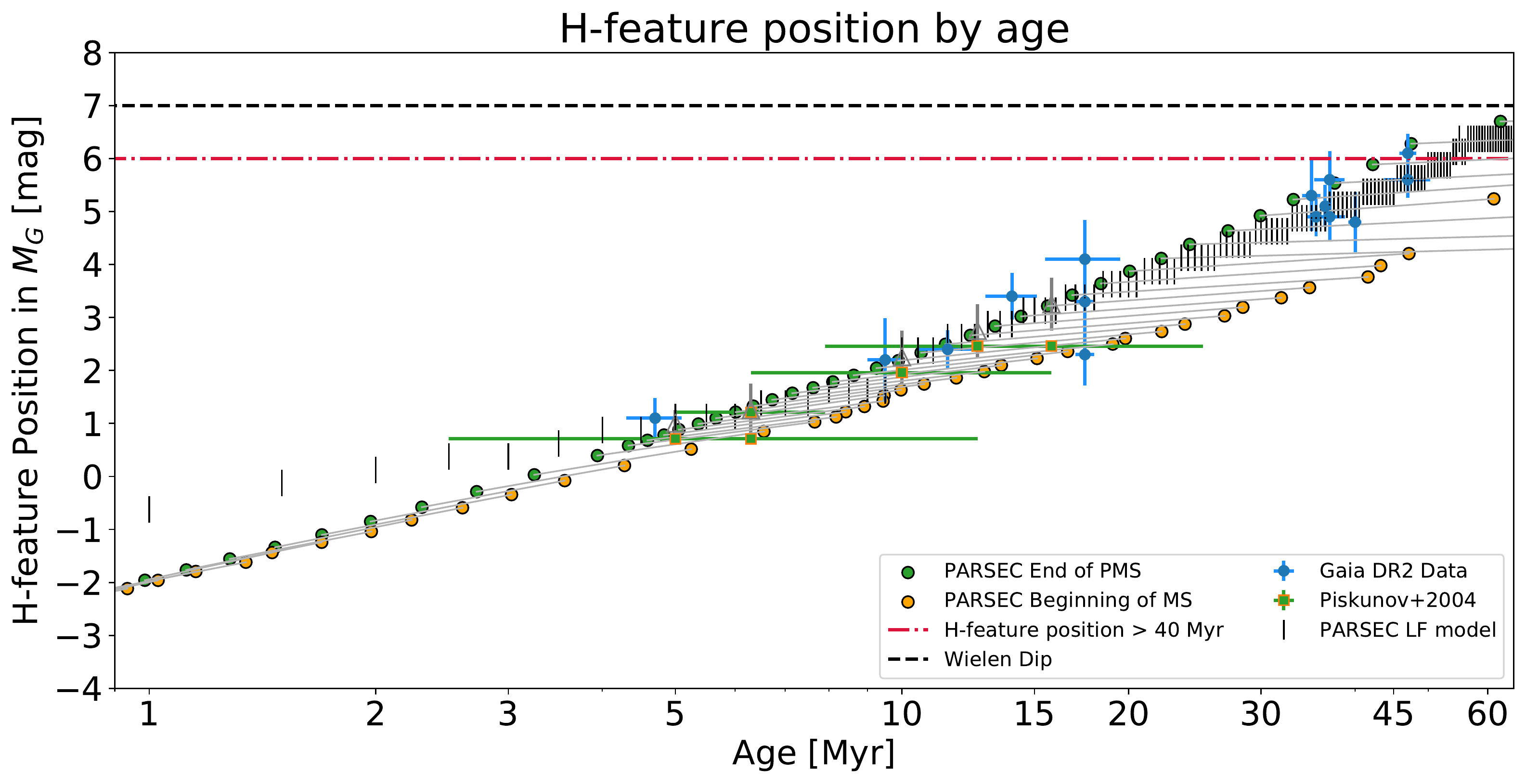}
                \caption{Relation between the age and the location of the H-peak is extracted from four sources.
                The location of the observed H-peak (blue points) were measured as a local maximum in Figure~\ref{fig:lumfunctionclusters} and \ref{fig:lumfunctionngc6231}. The error bars indicate the uncertainty range of the peak using the size of the kernel bandwidth as half of the error bar length, and the center dot indicates the position of the peak. 
                The green squares are   from \cite{Piskunov2004}. The black bars represent the range in magnitudes where the H-peak in the LF curve occurs as a local maximum, and the range of the bar is twice the bin size (0.25 mag). The green and yellow dots are critical points from the PARSEC evolutionary tracks, the relative location of the dots on the Hertzsprung--Russell diagram can be found in Fig.~\ref{fig:hrdparsec0}.}
                \label{fig:hloc-agerelation}
        \end{figure*}

        We plot the End of PMS and Beginning of MS points in Fig.~\ref{fig:hloc-agerelation} as green and yellow dots, respectively, and connect the points of the same initial mass with thin gray straight lines. The mass range shown is $\sim$0.7 -- 5 $M_\odot$, and some points are out of view. In general, the tracks show a trend that the lower the initial mass (luminosity), the later the End of PMS occurs, and the longer the transition phase lasts, which is represented by a longer gray line.

        Both sets of evolutionary points paint the outline of our observed H-peak--age relation. This is not a coincidence. As the stars in the H-peak section of the LF are believed to be in the transition phase from PMS to MS \citep{Piskunov1996}, the PARSEC evolutionary tracks clearly reflect this behavior. 
        
        \subsection{An empirical function for the time evolution of the H-feature}
        
        We have shown that observational and model data on the $M_G$-age plot in Fig.~\ref{fig:hloc-agerelation} are consistent. We can therefore derive the prediction function where a given age returns a certain range of the H-peak in $M_G$, and vice versa.
        
        The distribution of the data points indicates that we can fit a logarithmic relation, 
        \begin{equation}
                M_G^{\rm H-peak}(\tau) = a \log_{10}\left(\frac{\tau}{\rm yr}\right) + b
                \label{eq:MG_age}
        ,\end{equation}
        where $M_G^{\rm H-peak}$ is the position of the H-peak, $\tau$ is the age in yrs, and $a$ and $b$ are fitting parameters. By linear regression from 1 to 60 Myr in log-scale with the mean positions of the mass-luminosity relation predictions (black bars in Fig.~\ref{fig:hloc-agerelation}), we obtain \textbf{$(a_{\rm mean}, b_{\rm mean})  = (4.523; -28.921)$}. In order to estimate the uncertainties, we also perform a linear regression with the upper and lower points of the black bars, respectively, and eventually obtain three sets of parameters (see Table~\ref{tab:age-h_fit_param}).
        \begin{table}          
                \centering
                \caption{Fitting parameters for the function $M_G^{\rm H-peak}(\tau)$ using the LF predictions.}       
                \label{tab:age-h_fit_param}   
                \begin{threeparttable}
                            
                \begin{tabular}{lrr} 
                        \hline\\[-10pt]
                        Parameter   &$a$        &$b$\\[0pt]
                        \hline\\[-9pt]
                        Upper &4.523    &-28.921        \\
                        Mean  &4.523    &-29.171        \\
                        Lower &4.523    &-29.421        \\
                        \hline
                \end{tabular}
                \begin{tablenotes}
                        \item Note: The recommended applicable range of the function is $\tau$ $\in$ $(3, 30)$ Myr; the magnitude range $M_G^{\rm H-peak} \in (0, 4)$ mag. The entries ``Upper'' and ``Lower'' provide  bracketing relations for the uncertainties.
                \end{tablenotes}
                \end{threeparttable}  
        \end{table}            
        
        As we now have three functions predicting the mean, upper bound, and lower bound locations of the H-peak with a given age, we can use the inverse function to estimate the age $\tau$ of a population with a given location of the H-peak, as well as its upper bound $\tau_{\rm up}$ and lower bound $\tau_{\rm low}$. Representing Eq.~\ref{eq:MG_age} as $f(\tau)$ for simplicity of notation and its inverse as $g(\tau)$, we have
        \begin{eqnarray}
                \tau & = & g(M_G^{\rm H-peak}; a, b) = 10^{(M_G^{\rm H-peak} - b )/a}, \nonumber \\   
                \tau_{\rm low} & = & g(M_G^{\rm H-peak}; a_{\rm up}, b_{\rm up}),    \nonumber \\
                \tau_{\rm mean} & = & g(M_G^{\rm H-peak}; a_{\rm mean}, b_{\rm mean}),  \\
                \tau_{\rm up} & = & g(M_G^{\rm H-peak}; a_{\rm low}, b_{\rm low}). \nonumber
        \end{eqnarray}
        We note that the upper and lower bound of the $g$-function use, respectively, the lower and upper bound parameters from the $f$-function. Although the observed data allow the $M_G^{\rm H-peak}$ to range up to 6.2 mag, the effective domain of the functions should be limited to $M_G \in (-1, \sim4)$, as input beyond 4 mag returns an age with an uncertainty so large ($> 10$ Myr) that it becomes degenerate. 
        
\begin{figure*}
                \centering
                \includegraphics[width=1.0\linewidth]{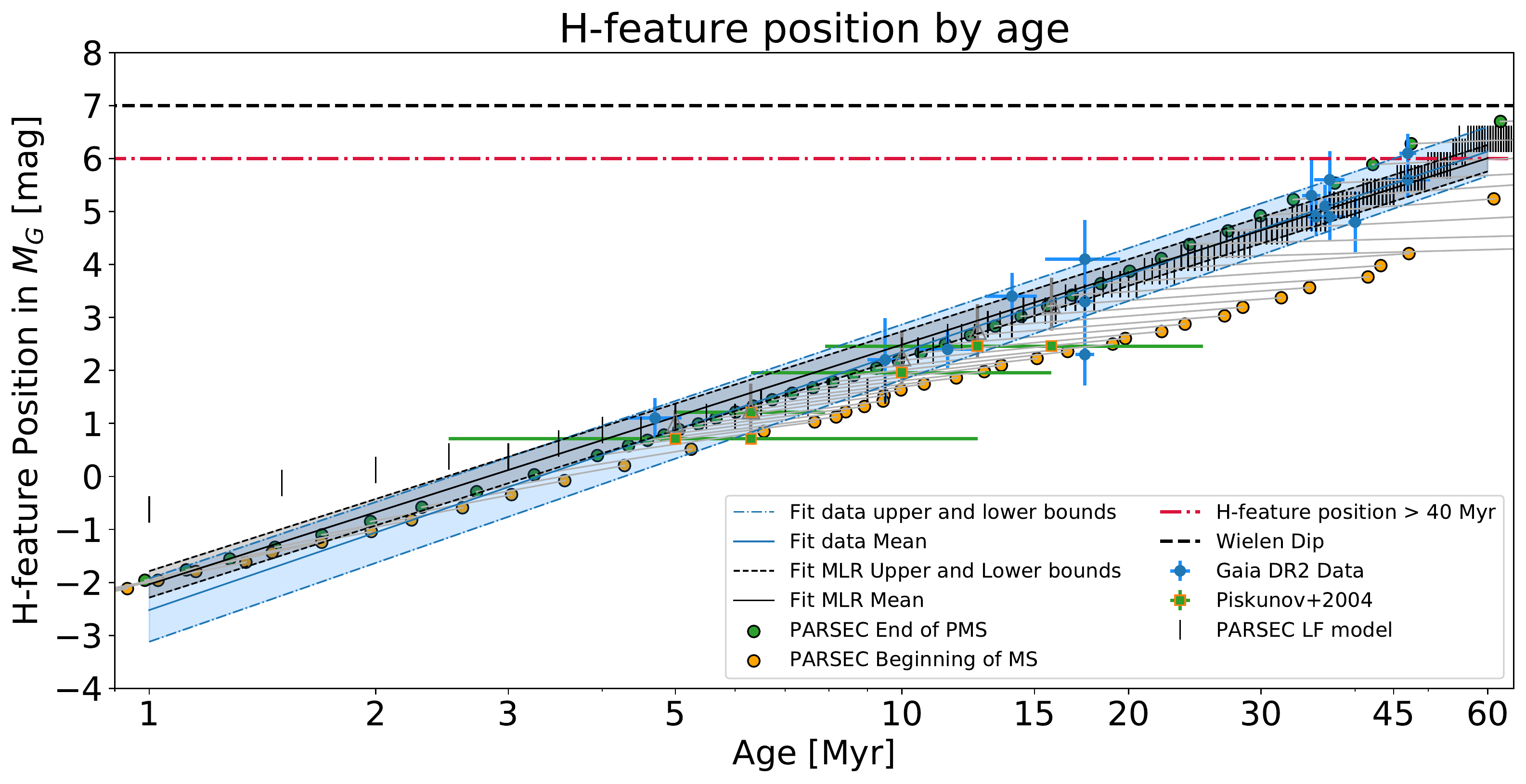}
                \caption{Fitted empirical age--H-peak relations. The data points are the same as in Fig.~\ref{fig:hloc-agerelation}. The blue solid line represents the peak-age function fitted from the observed data (blue points). The area shaded in blue is the uncertainty range, and the lower and upper bounds of the uncertainty are indicated by the thin blue dot-dashed lines. The black line is the peak-age function fitted from the PARSEC model points (black bars). The area shaded in gray is the uncertainty range and the lower and upper bounds are given by the thin black dashed lines.}
                \label{fig:hloc-agefit}
        \end{figure*}

        The fitted $f$-function (i.e., Eq.~\ref{eq:MG_age}) is plotted as a solid black line in Fig.~\ref{fig:hloc-agefit}, with its uncertainty limits as black dashed lines ($f_{\rm up}$ and $f_{\rm low}$). The space between the upper and lower bound functions are colored  light gray, indicating the estimated uncertainty of the prediction. All of our observational data points (in blue) lie within the gray area, or at least within the reach of their error bars. The fitted function runs up to 60 Myr, but the reader should take note that starting from 30 Myr, the location of the H-peak is stabilized at $M_G \simeq 6$ mag; consequently, it can no longer offer meaningful predictions beyond this age.
        
        We can use the same procedure to fit the \textit{Gaia} data points (in blue). The results are plotted as blue solid and dashed lines in Fig.~\ref{fig:hloc-agefit} and show a wider uncertainty range than the theoretical result based on the mass-luminosity relation (black solid line). The scarcity of data points for relatively young ages contributes to this uncertainty. Even so, the trend in the predicted and empirical relation are the same: the theoretical result is encompassed by the empirical relation in the full age domain. We list the detailed results of the alternative fit in Table~\ref{tab:age-h_data_fit_param} of Appendix~\ref{chap:LF|appendix:age-h_relation_data_fit}, as the focus of this work is the theoretical prediction of the H-peak--age relation.
        
        \section{Discussion}
        \label{chap:LF|sec:discussion}
        
        \subsection{Using the H-peak as an age indicator}
        The temporal behavior of the H-peak provides a simple method for estimating the age of young stellar populations ($< 30$ Myr) before the feature settles at $M_G \simeq 6$ mag for ages up to $\sim\,50-100$\,Myr. In this section we  discuss the limitations and intricacies of this chronometer. 
        
        Processes that may limit applicability include an overly small sample size (as a result of a limited number of stars present in the association or cluster) and extinction (which can reduce the effective resolution of the LF and of the effective sample size).
        Attenuation of starlight due to dust is particularly relevant for very young associations and clusters (with an age $< 5$ Myr) where newly formed PMS stars may still be embedded in their natal cloud environment. It may be particularly challenging to find such sufficiently populated young systems within a reasonable distance, though the case of the $4.7\pm0.4$ Myr old open cluster NGC 6231 (see Sect.~\ref{chap:LF|subsec:ngc6231}) shows that it is feasible. 
        A solution to heavy extinction is to observe the population at near-infrared wavelengths; for example, \cite{Belikov1997} showed that it is feasible in the K band.
        
        \subsection{The H-peak shifting due to metallicity}
        We assume solar metallicity (Z = 0.0152) for all the populations we study in this work, and choose the model's metallicity accordingly. However, as the luminosity and color of the stars at various evolutionary stages can be affected by metallicity, the position of the H-peak is shifted as well. Investigating the details of this effect is beyond the scope of this work; we present this discussion as a direction for future studies of the H-peak.
        
\begin{figure}[h]
                \centering
                \includegraphics[width=0.9\linewidth]{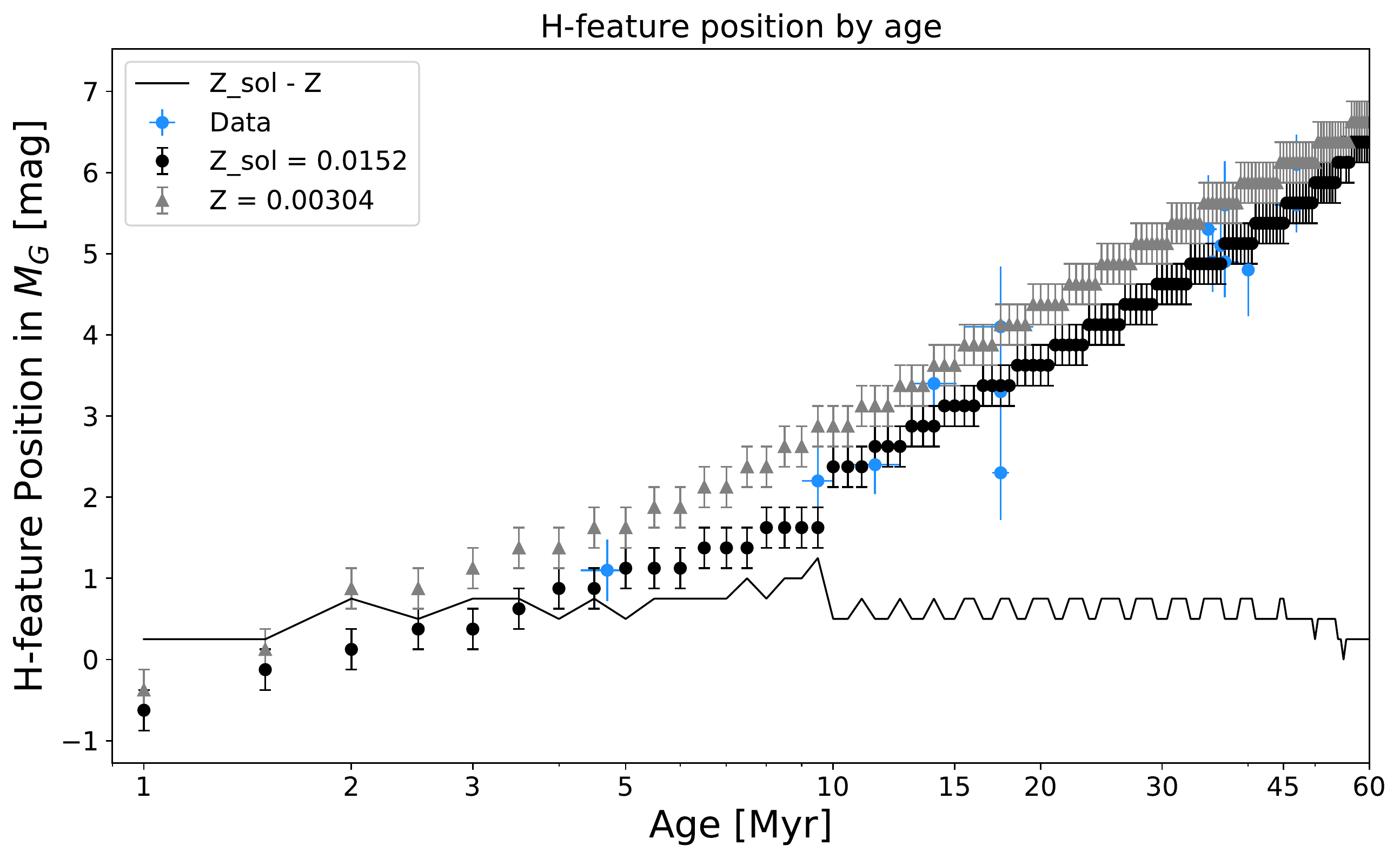}
                \caption{Predicted H-peak at lower metallicity has a lower brightness at a given time. The shift in the H-peak position can be explained by the shift in the evolutionary tracks with metallicity. As the metallicity becomes lower, the stars of the same initial mass become hotter and brighter. Therefore, given the same initial mass, the lower metallicity stars evolve faster on the pre-main sequence; at the same evolutionary stage, the critical points on the evolutionary tracks of lower metallicity represent a star with a lower mass and brightness. The lower metallicity population undergoes the critical phases earlier, shifting the H-peak--age relation to the lower age end.}
                \label{fig:hloc-age-metallicity}
        \end{figure}

        As an example, we compare the luminosity function of low metallicity (Z = 0.00304, one-fifth of solar metallicity), close to that of the Small Magellanic Cloud \citep[e.g.,][]{Mokiem2007} with that of solar metallicity in Fig.~\ref{fig:hloc-age-metallicity}. 
        At a given time, the H-peak at lower metallicity is shifted toward the faint end relative to higher metallicity. This originates from the systematic upward shift in both luminosity and temperature by a lower metallicity. As a result, given the same initial mass, the stars of lower metallicity evolve toward the ZAMS faster than those with a higher metallicity; in other words, the lower metallicity population undergoes the critical phases earlier, shifting the H-peak--age relation to the lower age end. Therefore, at a given time the population of lower metallicity would have a fainter H-peak in the luminosity function. This effect is further demonstrated and explained in Appendix~\ref{chap:LF|appendix:parsec_evotracks}.
        
        \subsection{The H-feature in the LF of the Solar neighborhood}
        
        The H-feature is a signature of populations much younger than the estimated average age of $> 10$ Gyr for the solar neighborhood \citep{Binney2000, Aumer2009}, and the PARSEC model predicts that it completely disappears after $\sim$ 200 Myr. If the solar environment were exclusively occupied by billion-year-old stars, the H-feature would not be observed, and as a consequence the Wielen dip next to the H-peak would lose contrast from the peak, becoming undetectable.
        The presence of the H-feature, at least in the Gliese and HIP2 samples, however, seems to point to a local population featuring a significant fraction of young and low-mass stars that are still in the phase of transitioning from PMS to MS. The existence of nearby large-scale loosely connected stellar streams such as the Pisces-Eridanus stream \citep[$\sim135$ Myr, see][]{Roser2020} suggests that lost members of these young structures can interlope into the solar environment. However, with the currently known closest young populations, the AB Doradus and Beta Pictoris moving groups \citep[e.g.,][]{Gagne2018}, we cannot yet confirm this hypothesis. A more detailed study of the structure of the LF in the solar neighborhood is needed to constrain the composition of the closest stars.

        \section{Conclusion}
        \label{chap:LF|sec:conclusion}
        In this work we investigated the fine structure of the luminosity functions in various populations using \textit{Gaia} DR2. We obtained the LF of the stars in the solar neighborhood within 20 pc, reproducing the result of \cite{Wielen1974} with the Wielen dip. We also produced the LF of 15 young stellar populations and identified the H-feature and the Wielen dip in all of them. 
        
        We took advantage of the high accuracy of the \textit{Gaia} data and state-of-the-art PARSEC models to demonstrate the origin and evolution of the H-feature using the mass-luminosity relation and its derivative curve, re-affirming that the phenomenon is observable. Our results supersede the calibration of \cite{Belikov1997}. 
        
        Finally, we fit an empirical function of the H-peak--age relation where, given the location of the H-peak in the LF of $M_G$, it is possible to  estimate the age of a population younger than 30 Myr without fitting the isochrones in the color-magnitude diagram. We note 
        that the H-peak observed in the population of solar neighborhood field stars may indicate that a significant fraction of this mixed population is younger than 200 Myr, but we are unable to confirm this with the known inventory of young stellar populations within the 20 pc distance limit; further detailed studies are needed to identify the age composition of the closest stars to the Sun.
        
        \section*{Acknowledgment}
        This work is financially supported by NOVA. 
        This work has made use of data from the European Space Agency (ESA) mission {\it Gaia}\footnote{\url{https://www.cosmos.esa.int/gaia}}, processed by the {\it Gaia} Data Processing and Analysis Consortium (DPAC)\footnote{\url{https://www.cosmos.esa.int/web/gaia/dpac/consortium}}. Funding for the DPAC has been provided by national institutions, in particular the institutions participating in the {\it Gaia} Multilateral Agreement.
        
        We thank an anonymous referee for a careful reading of the manuscript and dr. Phil Uttley for discussions on statistics.
        
        \bibliographystyle{aa}
        \bibliography{LF_in_Gaia}

\begin{thebibliography}{65}
\expandafter\ifx\csname natexlab\endcsname\relax\def\natexlab#1{#1}\fi

\bibitem[{Andrae {et~al.}(2018)Andrae, Fouesneau, Creevey, Ordenovic, Mary,
  Burlacu, Chaoul, Jean-Antoine-Piccolo, Kordopatis, Korn, Lebreton, Panem,
  Pichon, Th{\'{e}}venin, Walmsley, \& Bailer-Jones}]{Andrae2018}
Andrae, R., Fouesneau, M., Creevey, O., {et~al.} 2018, Astronomy {\&}
  Astrophysics, 616, A8

\bibitem[{Aumer \& Binney(2009)}]{Aumer2009}
Aumer, M. \& Binney, J.~J. 2009, MNRAS, 397, 1286

\bibitem[{Bailer-Jones(2015)}]{Bailer-Jones2015}
Bailer-Jones, C. A.~L. 2015, Publications of the Astronomical Society of the
  Pacific, 127, 994

\bibitem[{Bate(2012)}]{Bate2012}
Bate, M.~R. 2012, Monthly Notices of the Royal Astronomical Society, Volume
  419, Issue 4, pp. 3115-3146., 419, 3115

\bibitem[{Belikov(1997)}]{Belikov1997}
Belikov, A.~N. 1997, Astronomical {\&} Astrophysical Transactions, 14, 19

\bibitem[{Belikov {et~al.}(1998)Belikov, Hirte, Meusinger, Piskunov, \&
  Schilbach}]{Belikov1998}
Belikov, A.~N., Hirte, S., Meusinger, H., Piskunov, A.~E., \& Schilbach, E.
  1998, ASTRONOMY AND ASTROPHYSICS, 332, 575

\bibitem[{{Belikov} \& {Piskunov}(1997)}]{BelikovEtPiskunov1997}
{Belikov}, A.~N. \& {Piskunov}, A.~E. 1997, Astronomy Reports, 41, 28

\bibitem[{Bessell \& Stringfellow(1993)}]{Bessell1993}
Bessell, M.~S. \& Stringfellow, G.~S. 1993, Annual Review of Astronomy and
  Astrophysics, 31, 433

\bibitem[{Binks {et~al.}(2020)Binks, Jeffries, \& Wright}]{Binks2020}
Binks, A.~S., Jeffries, R.~D., \& Wright, N.~J. 2020, Monthly Notices of the
  Royal Astronomical Society, 494, 2429

\bibitem[{Binney {et~al.}(2000)Binney, Dehnen, \& Bertelli}]{Binney2000}
Binney, J., Dehnen, W., \& Bertelli, G. 2000, MNRAS, 318, 658

\bibitem[{Bossini {et~al.}(2019)Bossini, Vallenari, Bragaglia, Cantat-Gaudin,
  Sordo, Balaguer-N{\'{u}}{\~{n}}ez, Jordi, Moitinho, Soubiran, Casamiquela,
  Carrera, \& Heiter}]{Bossini2019}
Bossini, D., Vallenari, A., Bragaglia, A., {et~al.} 2019, Astronomy {\&}
  Astrophysics, 623, A108

\bibitem[{Boubert \& Everall(2020)}]{Boubert2020}
Boubert, D. \& Everall, A. 2020, Monthly Notices of the Royal Astronomical
  Society, 497, 4246

\bibitem[{Bouret {et~al.}(2015)Bouret, Lanz, Hillier, Martins, Marcolino, \&
  Depagne}]{Bouret2015}
Bouret, J.-C., Lanz, T., Hillier, D.~J., {et~al.} 2015, Monthly Notices of the
  Royal Astronomical Society, 449, 1545

\bibitem[{Bouy \& Alves(2015)}]{Bouy2015}
Bouy, H. \& Alves, J. 2015, Astronomy {\&} Astrophysics, 584, A26

\bibitem[{Bressan {et~al.}(2012)Bressan, Marigo, Girardi, Salasnich, Cero,
  Rubele, Nanni, Bressan, Marigo, Girardi, Salasnich, {Dal Cero}, Rubele, \&
  Nanni}]{Bressan2012}
Bressan, A., Marigo, P., Girardi, L., {et~al.} 2012, MNRAS, 427, 127

\bibitem[{Cantat-Gaudin {et~al.}(2018)Cantat-Gaudin, Jordi, Vallenari,
  Bragaglia, Balaguer-N{\'{u}}{\~{n}}ez, Soubiran, Bossini, Moitinho,
  Castro-Ginard, Krone-Martins, Casamiquela, Sordo, \&
  Carrera}]{Cantat-Gaudin2018b}
Cantat-Gaudin, T., Jordi, C., Vallenari, A., {et~al.} 2018, Astronomy {\&}
  Astrophysics, 618, A93

\bibitem[{Cantat-Gaudin {et~al.}(2019{\natexlab{a}})Cantat-Gaudin, Jordi,
  Wright, Armstrong, Vallenari, Balaguer-N{\'{u}}{\~{n}}ez, Ramos, Bossini,
  Padoan, Pelkonen, Mapelli, \& Jeffries}]{Cantat-Gaudin2019b}
Cantat-Gaudin, T., Jordi, C., Wright, N.~J., {et~al.} 2019{\natexlab{a}},
  Astronomy {\&} Astrophysics, 626, A17

\bibitem[{Cantat-Gaudin {et~al.}(2019{\natexlab{b}})Cantat-Gaudin, Mapelli,
  Balaguer-N{\'{u}}{\~{n}}ez, Jordi, Sacco, \& Vallenari}]{Cantat-Gaudin2019a}
Cantat-Gaudin, T., Mapelli, M., Balaguer-N{\'{u}}{\~{n}}ez, L., {et~al.}
  2019{\natexlab{b}}, Astronomy {\&} Astrophysics, 621, A115

\bibitem[{Chabrier(2001)}]{Chabrier2001}
Chabrier, G. 2001, The Astrophysical Journal, 554, 1274

\bibitem[{Chen {et~al.}(2015)Chen, Bressan, Girardi, Marigo, Kong, Lanza, Chen,
  Bressan, Girardi, Marigo, Kong, \& Lanza}]{Chen2015}
Chen, Y., Bressan, A., Girardi, L., {et~al.} 2015, MNRAS, 452, 1068

\bibitem[{Chen {et~al.}(2014)Chen, Girardi, Bressan, Marigo, Barbieri, Kong,
  Chen, Girardi, Bressan, Marigo, Barbieri, \& Kong}]{Chen2014}
Chen, Y., Girardi, L., Bressan, A., {et~al.} 2014, MNRAS, 444, 2525

\bibitem[{Chen {et~al.}(2019)Chen, Girardi, Fu, Bressan, Aringer, {Dal Tio},
  Pastorelli, Marigo, Costa, \& Zhang}]{Chen2019}
Chen, Y., Girardi, L., Fu, X., {et~al.} 2019, Astronomy {\&} Astrophysics, 632,
  A105

\bibitem[{Dantona \& Mazzitelli(1985)}]{Dantona1985}
Dantona, F. \& Mazzitelli, I. 1985, The Astrophysical Journal, 296, 502

\bibitem[{Evans {et~al.}(2018)Evans, Riello, {De Angeli}, Carrasco,
  Montegriffo, Fabricius, Jordi, Palaversa, Diener, Busso, Cacciari, \& van
  Leeuwen}]{Evans2018}
Evans, D.~W., Riello, M., {De Angeli}, F., {et~al.} 2018, Astronomy {\&}
  Astrophysics, 616, A4

\bibitem[{Gagn{\'{e}} {et~al.}(2020)Gagn{\'{e}}, David, Mamajek, Mann, Faherty,
  \& B{\'{e}}dard}]{Gagne2020}
Gagn{\'{e}}, J., David, T.~J., Mamajek, E.~E., {et~al.} 2020, The Astrophysical
  Journal, 903, 96

\bibitem[{Gagn{\'{e}} {et~al.}(2018)Gagn{\'{e}}, Mamajek, Malo, Riedel,
  Rodriguez, Lafreni{\`{e}}re, Faherty, Roy-Loubier, Pueyo, Robin, \&
  Doyon}]{Gagne2018}
Gagn{\'{e}}, J., Mamajek, E.~E., Malo, L., {et~al.} 2018, The Astrophysical
  Journal, 856, 23

\bibitem[{{Gaia Collaboration}(2016)}]{GaiaCollaboration2016}
{Gaia Collaboration}. 2016, Astronomy {\&} Astrophysics, 595, A1

\bibitem[{{Gaia Collaboration} {et~al.}(2018){Gaia Collaboration}, Brown,
  Vallenari, Prusti, de~Bruijne, Babusiaux, \&
  Bailer-Jones}]{GaiaCollaboration2018a}
{Gaia Collaboration}, Brown, A. G.~A., Vallenari, A., {et~al.} 2018, Astronomy
  {\&} Astrophysics, 616, A1

\bibitem[{Gliese(1969)}]{Gliese1969}
Gliese, W. 1969, Veroeffentlichungen des Astronomischen Rechen-Instituts
  Heidelberg, 22, 1

\bibitem[{Gliese(2015)}]{Gliese2015}
Gliese, W. 2015, VizieR Online Data Catalog, V/1

\bibitem[{{Gliese} \& {Jahrei{\ss}}(1991)}]{Gliese1991}
{Gliese}, W. \& {Jahrei{\ss}}, H. 1991, {Preliminary Version of the Third
  Catalogue of Nearby Stars}, On: The Astronomical Data Center CD-ROM: Selected
  Astronomical Catalogs

\bibitem[{Gliese \& Jahreiss(1995)}]{Gliese1995}
Gliese, W. \& Jahreiss, H. 1995, VizieR Online Data Catalog, V/70A

\bibitem[{Haywood(1994)}]{Haywood1994}
Haywood, M. 1994, Astronomy and Astrophysics (ISSN 0004-6361), vol. 282, no. 2,
  p. 444-451, 282, 444

\bibitem[{Hinkel {et~al.}(2017)Hinkel, Mamajek, Turnbull, Osby, Shkolnik,
  Smith, Klimasewski, Somers, \& Desch}]{Hinkel2017}
Hinkel, N.~R., Mamajek, E.~E., Turnbull, M.~C., {et~al.} 2017, The
  Astrophysical Journal, 848, 34

\bibitem[{{Iben, Icko}(1965)}]{Iben1965}
{Iben, Icko}, J. 1965, The Astrophysical Journal, 141, 993

\bibitem[{Jao {et~al.}(2018)Jao, Henry, Gies, \& Hambly}]{Jao2018}
Jao, W.-C., Henry, T.~J., Gies, D.~R., \& Hambly, N.~C. 2018, The Astrophysical
  Journal, 861, L11

\bibitem[{Jeffries {et~al.}(2001)Jeffries, Thurston, \& Hambly}]{Jeffries2001}
Jeffries, R.~D., Thurston, M.~R., \& Hambly, N.~C. 2001, Astronomy {\&}
  Astrophysics, 375, 863

\bibitem[{J{\o}rgensen \& Lindegren(2005)}]{Joergensen2005}
J{\o}rgensen, B.~R. \& Lindegren, L. 2005, Astronomy {\&} Astrophysics, 436,
  127

\bibitem[{Kroupa(2002)}]{Kroupa2002}
Kroupa, P. 2002, Science (New York, N.Y.), 295, 82

\bibitem[{Kroupa {et~al.}(1990)Kroupa, Tout, \& Gilmore}]{Kroupa1990}
Kroupa, P., Tout, C.~A., \& Gilmore, G. 1990, Monthly Notices of the Royal
  Astronomical Society, 244, 76

\bibitem[{Lee(1988)}]{Lee1988}
Lee, S.-W. 1988, Vistas in Astronomy, 31, 445

\bibitem[{Lee \& Sung(1995)}]{Lee1995}
Lee, S.-W. \& Sung, H. 1995, Journal of the Korean Astronomical Society, vol.
  28, no. 1, p. 45-59, 28, 45

\bibitem[{Lindegren(2018)}]{LL:LL-124}
Lindegren, L. 2018, {Re-normalising the astrometric chi-square in \textit{Gaia}
  DR2}

\bibitem[{Lindegren {et~al.}(2018)Lindegren, Hern{\'{a}}ndez, Bombrun, Klioner,
  Bastian, Ramos-Lerate, de~Torres, Steidelm{\"{u}}ller, Stephenson, Hobbs,
  Lammers, Biermann, Geyer, Hilger, Michalik, Stampa, McMillan,
  Casta{\~{n}}eda, Clotet, Comoretto, Davidson, Fabricius, Gracia, Hambly,
  Hutton, Mora, Portell, van Leeuwen, Abbas, Abreu, Altmann, Andrei, Anglada,
  Balaguer-N{\'{u}}{\~{n}}ez, Barache, Becciani, Bertone, Bianchi, Bouquillon,
  Bourda, Br{\"{u}}semeister, Bucciarelli, Busonero, Buzzi, Cancelliere,
  Carlucci, Charlot, Cheek, Crosta, Crowley, de~Bruijne, de~Felice, Drimmel,
  Esquej, Fienga, Fraile, Gai, Garralda, Gonz{\'{a}}lez-Vidal, Guerra, Hauser,
  Hofmann, Holl, Jordan, Lattanzi, Lenhardt, Liao, Licata, Lister,
  L{\"{o}}ffler, Marchant, Martin-Fleitas, Messineo, Mignard, Morbidelli,
  Poggio, Riva, Rowell, Salguero, Sarasso, Sciacca, Siddiqui, Smart, Spagna,
  Steele, Taris, Torra, van Elteren, van Reeven, \& Vecchiato}]{Lindegren2018}
Lindegren, L., Hern{\'{a}}ndez, J., Bombrun, A., {et~al.} 2018, Astronomy {\&}
  Astrophysics, 616, A2

\bibitem[{Luyten(1968)}]{Luyten1968}
Luyten, W.~J. 1968, Monthly Notices of the Royal Astronomical Society, 139, 221

\bibitem[{{Ma{\'{i}}z Apell{\'{a}}niz} \& Weiler(2018)}]{Apellaniz2018}
{Ma{\'{i}}z Apell{\'{a}}niz}, J. \& Weiler, M. 2018, Astronomy {\&}
  Astrophysics, 619, A180

\bibitem[{Marigo {et~al.}(2017)Marigo, Girardi, Bressan, Rosenfield, Aringer,
  Chen, Dussin, Nanni, Pastorelli, Rodrigues, Trabucchi, Bladh, Dalcanton,
  Groenewegen, Montalb{\'{a}}n, \& Wood}]{Marigo2017}
Marigo, P., Girardi, L., Bressan, A., {et~al.} 2017, The Astrophysical Journal,
  835, 77

\bibitem[{Marrese {et~al.}(2018)Marrese, Marinoni, Fabrizio, \&
  Altavilla}]{Marrese2018}
Marrese, P.~M., Marinoni, S., Fabrizio, M., \& Altavilla, G. 2018, Astronomy
  {\&} Astrophysics, 621, A144

\bibitem[{Mokiem {et~al.}(2007)Mokiem, de~Koter, Vink, Puls, Evans, Smartt,
  Crowther, Herrero, Langer, Lennon, Najarro, \& Villamariz}]{Mokiem2007}
Mokiem, M.~R., de~Koter, A., Vink, J.~S., {et~al.} 2007, Astronomy {\&}
  Astrophysics, 473, 603

\bibitem[{Naylor {et~al.}(2002)Naylor, Totten, Jeffries, Pozzo, Devey, \&
  Thompson}]{Naylor2002}
Naylor, T., Totten, E.~J., Jeffries, R.~D., {et~al.} 2002, Monthly Notices of
  the Royal Astronomical Society, Volume 335, Issue 2, pp. 291-310., 335, 291

\bibitem[{Olivares {et~al.}(2019)Olivares, Bouy, Sarro, Miret-Roig, Berihuete,
  Bertin, Barrado, Hu{\'{e}}lamo, Tamura, Allen, Beletsky, Serre, \&
  Cuillandre}]{Olivares2019}
Olivares, J., Bouy, H., Sarro, L.~M., {et~al.} 2019, Astronomy {\&}
  Astrophysics, 625, A115

\bibitem[{Palla \& Stahler(1993)}]{Palla1993}
Palla, F. \& Stahler, S.~W. 1993, The Astrophysical Journal, 418, 414

\bibitem[{Pecaut \& Mamajek(2013)}]{Pecaut2013}
Pecaut, M.~J. \& Mamajek, E.~E. 2013, The Astrophysical Journal Supplement
  Series, 208, 9

\bibitem[{Pecaut \& Mamajek(2016)}]{Pecaut2016}
Pecaut, M.~J. \& Mamajek, E.~E. 2016, Monthly Notices of the Royal Astronomical
  Society, 461, 794

\bibitem[{Perryman {et~al.}(1997)Perryman, {European Space Agency.}, \& {FAST
  Consortium.}}]{Perryman1997}
Perryman, M. A.~C., {European Space Agency.}, \& {FAST Consortium.} 1997, {The
  Hipparcos and Tycho catalogues : astrometric and photometric star catalogues
  derived from the ESA Hipparcos Space Astrometry Mission} (ESA Publications
  Division)

\bibitem[{Piskunov \& Belikov(1996)}]{Piskunov1996}
Piskunov, A. \& Belikov, A. 1996, Astronomy Letters, 22, 466

\bibitem[{Piskunov(2001)}]{Piskunov2001}
Piskunov, A.~E. 2001, Bulletin of the Astronomical Society of India, 29, 259

\bibitem[{Piskunov {et~al.}(2004)Piskunov, Belikov, Kharchenko, Sagar,
  Subramaniam, Piskunov, Belikov, Kharchenko, Sagar, \&
  Subramaniam}]{Piskunov2004}
Piskunov, A.~E., Belikov, A.~N., Kharchenko, N.~V., {et~al.} 2004, MNRAS, 349,
  1449

\bibitem[{R{\"{o}}ser \& Schilbach(2020)}]{Roser2020}
R{\"{o}}ser, S. \& Schilbach, E. 2020, Astronomy {\&} Astrophysics, 638, A9

\bibitem[{Salpeter(1955)}]{Salpeter1955}
Salpeter, E.~E. 1955, The Astrophysical Journal, 121, 161

\bibitem[{Stauffer {et~al.}(2010)Stauffer, Tanner, Bryden, Ramirez, Berriman,
  Ciardi, Kane, Mizusawa, Payne, Plavchan, von Braun, Wyatt, \&
  Kirkpatrick}]{Stauffer2010}
Stauffer, J., Tanner, A.~M., Bryden, G., {et~al.} 2010, Publications of the
  Astronomical Society of the Pacific, 122, 885

\bibitem[{Sung {et~al.}(2013)Sung, Sana, \& Bessell}]{Sung2013}
Sung, H., Sana, H., \& Bessell, M.~S. 2013, The Astronomical Journal, 145, 37

\bibitem[{Tang {et~al.}(2014)Tang, Bressan, Rosenfield, Slemer, Marigo,
  Girardi, \& Bianchi}]{Tang2014}
Tang, J., Bressan, A., Rosenfield, P., {et~al.} 2014, Monthly Notices of the
  Royal Astronomical Society, 445, 4287

\bibitem[{van Leeuwen(2007)}]{VanLeeuwen2007}
van Leeuwen, F. 2007, Astrophysics and Space Science Library, Vol. 350,
  {Hipparcos, the New Reduction of the Raw Data}, ed. F.~van Leeuwen
  (Dordrecht: Springer Netherlands)

\bibitem[{Wielen(1974)}]{Wielen1974}
Wielen, R. 1974, In: Highlights of astronomy. Volume 3. (A75-21577 08-88)
  Dordrecht, D. Reidel Publishing Co., 1974, p. 395-407; Discussion, p. 407.,
  3, 395

\end{thebibliography}
        
        \appendix
        \section{Selection of objects in the solar vicinity}
        
        \subsection{Data from \textit{Gaia} DR2}
        \label{chap:LF|appendix:data_select_gaia_dr2}
        The code below selects all objects within 20 pc of the Sun. \footnote{The names of the columns and their contents can be found at \url{https://gea.esac.esa.int/archive/documentation/GDR2/Gaia_archive/chap_datamodel/sec_dm_main_tables/ssec_dm_gaia_source.html}}
        
        After downloading the data, we further trimmed down the data set with additional filters. First is the RUWE filter described in the technical note of \cite{LL:LL-124} for good quality   astrometry (RUWE $< 1.4$), then the BP/RP excess filter from \cite{Evans2018} to exclude sources with suspect photometry (\texttt{phot\_bp\_rp\_excess\_factor}~$< (1.3 + 0.06 (BP - RP)^2)$).
        
\begin{verbatim}
SELECT gaia.designation,
gaia.phot_g_mean_mag, gaia.phot_g_mean_flux,
gaia.phot_g_mean_flux_error,
gaia.phot_bp_rp_excess_factor,
gaia.phot_bp_mean_mag, gaia.phot_bp_mean_flux,
gaia.phot_bp_mean_flux_error,
gaia.phot_rp_mean_mag, gaia.phot_rp_mean_flux,
gaia.phot_rp_mean_flux_error,
gaia.bp_rp,

gaia.parallax, gaia.parallax_error,
gaia.parallax_over_error,
ruwe.ruwe, gaia.l, gaia.b, gaia.ra, gaia.dec,
gaia.pmra, gaia.pmdec,

FROM gaiadr2.gaia_source AS gaia

LEFT OUTER JOIN gaiadr2.ruwe as ruwe
ON (gaia.source_id = ruwe.source_id)

WHERE (gaia.parallax >= 50)
\end{verbatim}
        
        \subsection{Cross-matching the \textit{Hipparcos} catalog to \textit{Gaia} DR2}
        We use the improved \textit{Hipparcos} to \textit{Gaia} DR2 cross-match table listed on the ``Known issues of \textit{Gaia} DR2'' web page to identify \textit{Hipparcos} objects in \textit{Gaia} DR2 .
        
        In this cross-match we first upload the \textit{Hipparcos}--\textit{Gaia} DR2 cross-match table by \cite{Marrese2018} as a user table (named \texttt{user\_dguo.hip2gdr2crossmatch}) into the author's account of \textit{Gaia} Archive. The user should upload this table to their own \textit{Gaia} Archive account and change the table name in the code below accordingly. 
        
\begin{verbatim}
SELECT gaia.designation,
gaia.phot_g_mean_mag, 
gaia.phot_g_mean_flux, 
gaia.phot_g_mean_flux_error, 
gaia.phot_bp_rp_excess_factor,
gaia.phot_bp_mean_mag, 
gaia.phot_bp_mean_flux, 
gaia.phot_bp_mean_flux_error,
gaia.phot_rp_mean_mag, 
gaia.phot_rp_mean_flux, 
gaia.phot_rp_mean_flux_error,  

gaia.parallax, gaia.parallax_error, gaia.bp_rp, 
ruwe.ruwe, gaia.l, gaia.b, gaia.ra, gaia.dec, 
gaia.pmra, gaia.pmdec, 
gaia.radial_velocity, 
gaia.radial_velocity_error,
dist.r_est, dist.r_lo, dist.r_hi, dist.r_len, 
gaia.teff_val, gaia.lum_val, 
a_g_val, e_bp_min_rp_val, hip2xmatch.hip

FROM user_dguo.hip2gdr2crossmatch AS hip2xmatch

LEFT OUTER JOIN 
gaiadr2.gaia_source AS gaia
ON (gaia.source_id = hip2xmatch.gdr2_source_id)

LEFT OUTER JOIN 
gaiadr2.ruwe AS ruwe
ON (gaia.source_id = ruwe.source_id)

LEFT OUTER JOIN 
external.gaiadr2_geometric_distance AS dist
ON (gaia.source_id = dist.source_id)
        
\end{verbatim}

        \section{Empirical function of the Age--H-peak position relation by fitting with \textit{Gaia} DR2 data}
        \label{chap:LF|appendix:age-h_relation_data_fit}
        In Table~\ref{tab:age-h_data_fit_param} we list the results of fitting the \textit{Gaia} data to the H-peak--age relation. The meaning of the parameters are as in Table~\ref{tab:age-h_fit_param}.
        
        \begin{table}[h!]        
                \centering
                \caption{Fitting parameters for the function $f(\tau)$ with \textit{Gaia} observations, where $\tau$ denotes age.}       
                \label{tab:age-h_data_fit_param}     
                \begin{threeparttable}
                    
                \begin{tabular}{lrr} 
                        \hline
                        Parameter   &$a$        &$b$\\
                        \hline
                        Upper   &4.791  &-30.672        \\
                        Mean    &4.867  &-31.725        \\
                        Lower   &4.943  &-32.778        \\
                        \hline
                \end{tabular}
                \begin{tablenotes}
                        \item Note: The effective applicable range of the function is $\tau$ $\in$ [1, 30] Myr, and the range of output $M_G \in (-3, 5)$ mag.
                \end{tablenotes}
                \end{threeparttable}
        \end{table}

        \section{The HRD of PARSEC evolutionary tracks}
        \label{chap:LF|appendix:parsec_evotracks}
        
\begin{figure}
                \centering
                \includegraphics[width=1.0\linewidth]{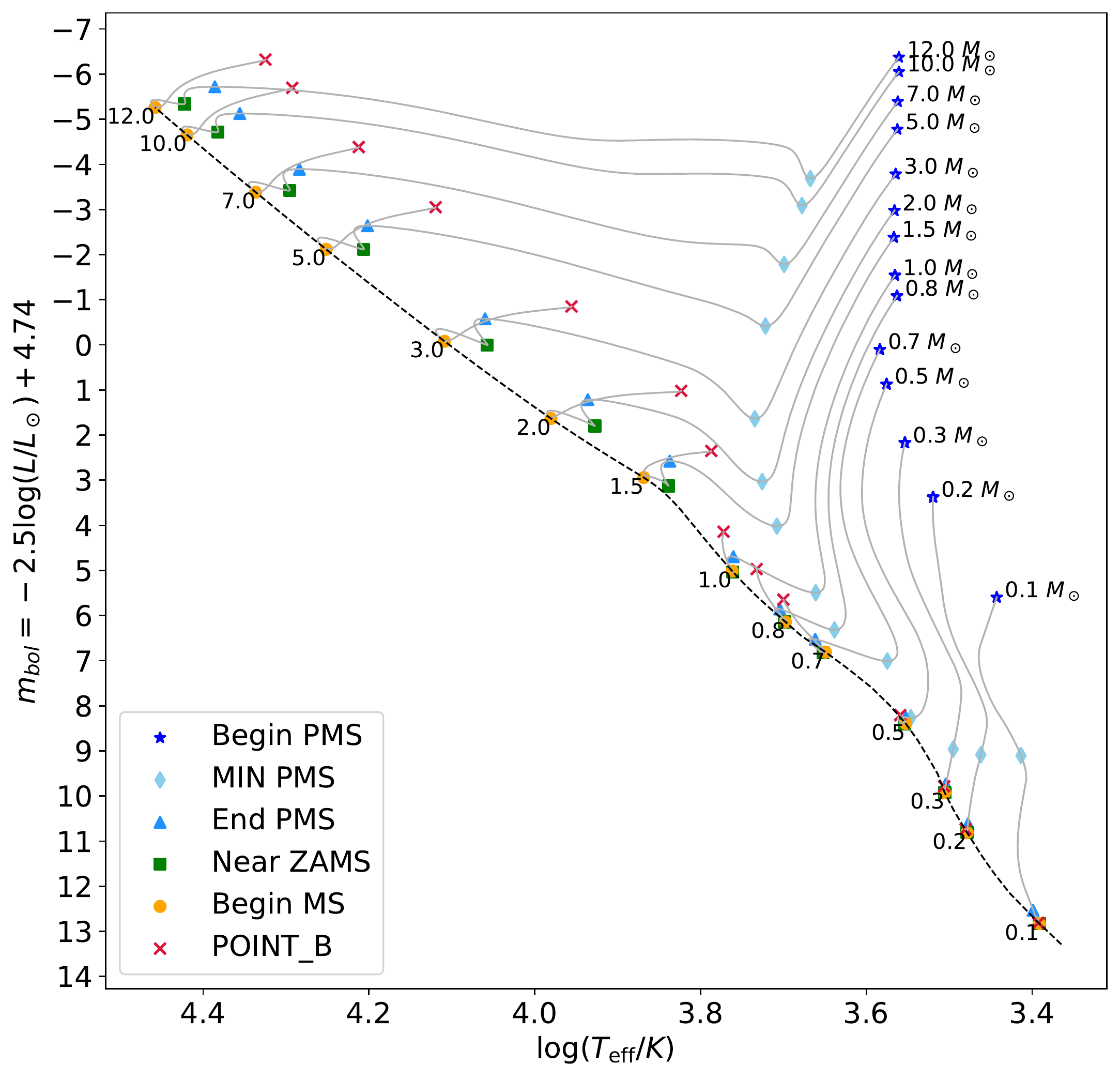}
                \caption{PARSEC evolutionary tracks of Z = 0.014 (to represent the solar metallicity) from the phase of Beginning of PMS (t=0, blue stars) to Point B (red crosses), up to an initial mass of 12 $M_\odot$. The gray lines are the evolutionary tracks, and the symbols indicate the critical phases along the track defined by \citet{Bressan2012}. All the points of Begin MS are connected by a dashed line to represent the ZAMS.}
                \label{fig:hrdparsec0}
        \end{figure}

        Figure~\ref{fig:hrdparsec0} shows the HRD including the PARSEC evolutionary tracks of initial mass from 0.1 $M_\odot$ up to 10 $M_\odot$. The diagram shows the critical points in various symbols. These points are defined in the \texttt{readme} file \footnote{\url{https://people.sissa.it/~sbressan/CAF09_V1.2S_M36_LT/readme.txt}} of the evolutionary tracks by \cite{Bressan2012}.
        
\begin{figure}
                \centering
                \includegraphics[width=1.0\linewidth]{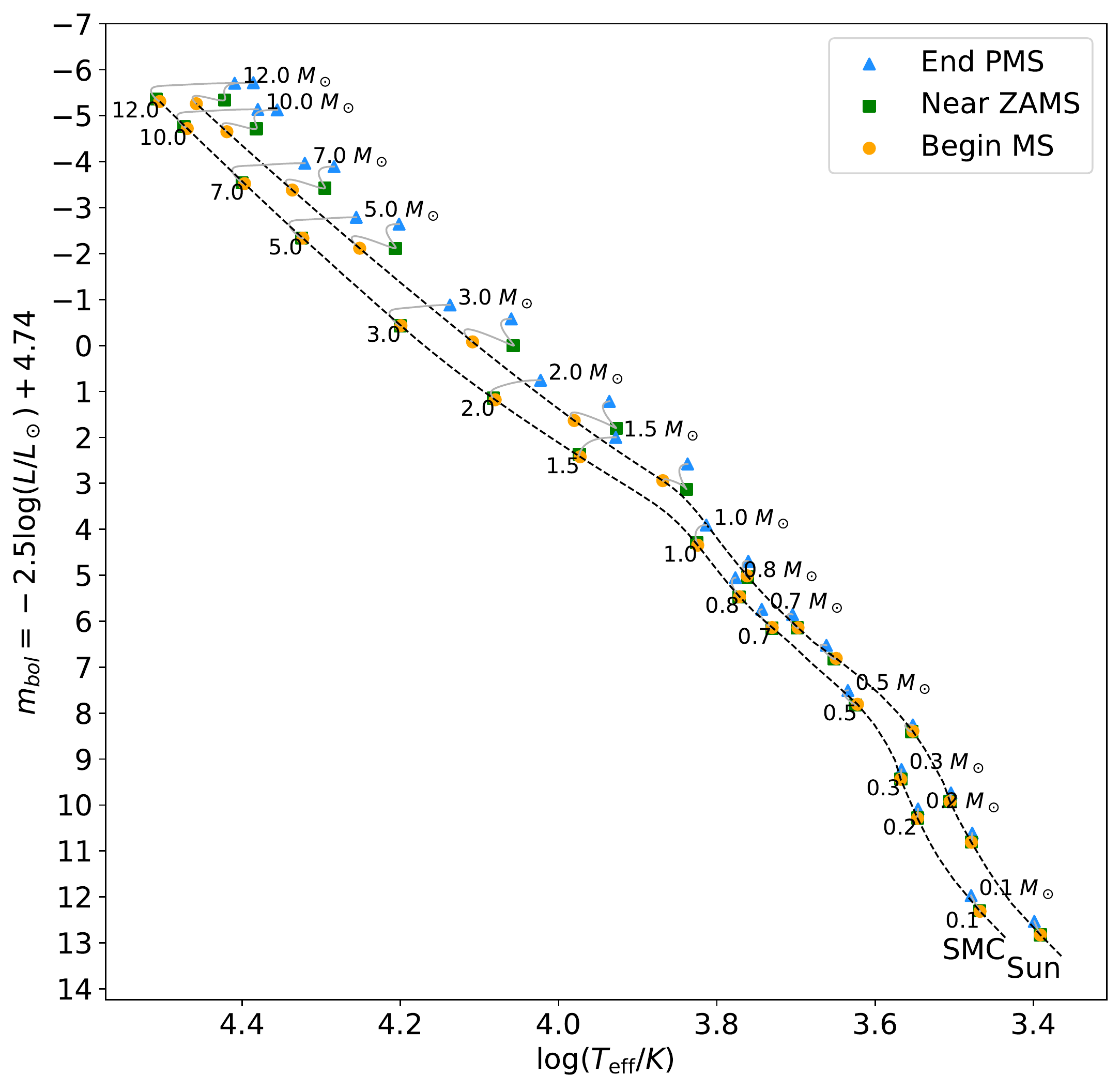}
                \caption{PARSEC evolutionary tracks of both Z = 0.014 and Z = 0.002, truncated at the phases between End of PMS and Begin MS (ZAMS). The tracks of Z = 0.002 represent the SMC metallicity Z $\simeq 1/5$~Z$_\odot$. The tracks of lower Z shift towards higher temperature and higher luminosity. Thus, for each pair of the tracks with the same initial mass, the ones  on the left have Z = 0.002. The End of PMS and Begin MS points on the Z = 0.002 tracks are labeled with the value of initial mass.}
                \label{fig:HRDparsec_zams}
        \end{figure}

        Figure~\ref{fig:HRDparsec_zams} presents both sets of tracks of Z = 0.002 (to represent the SMC's metallicity) and Z = 0.014 (to represent the Sun's metallicity), but truncated to show only the phases between End PMS and Begin MS. This diagram shows that for the same initial mass, the lower metallicity makes the stars brighter and hotter. However, what it does not show is the evolutionary speed and the time it takes a PMS star to reach the phase End PMS.
        
\begin{figure}
                \centering
                \includegraphics[width=0.9\linewidth]{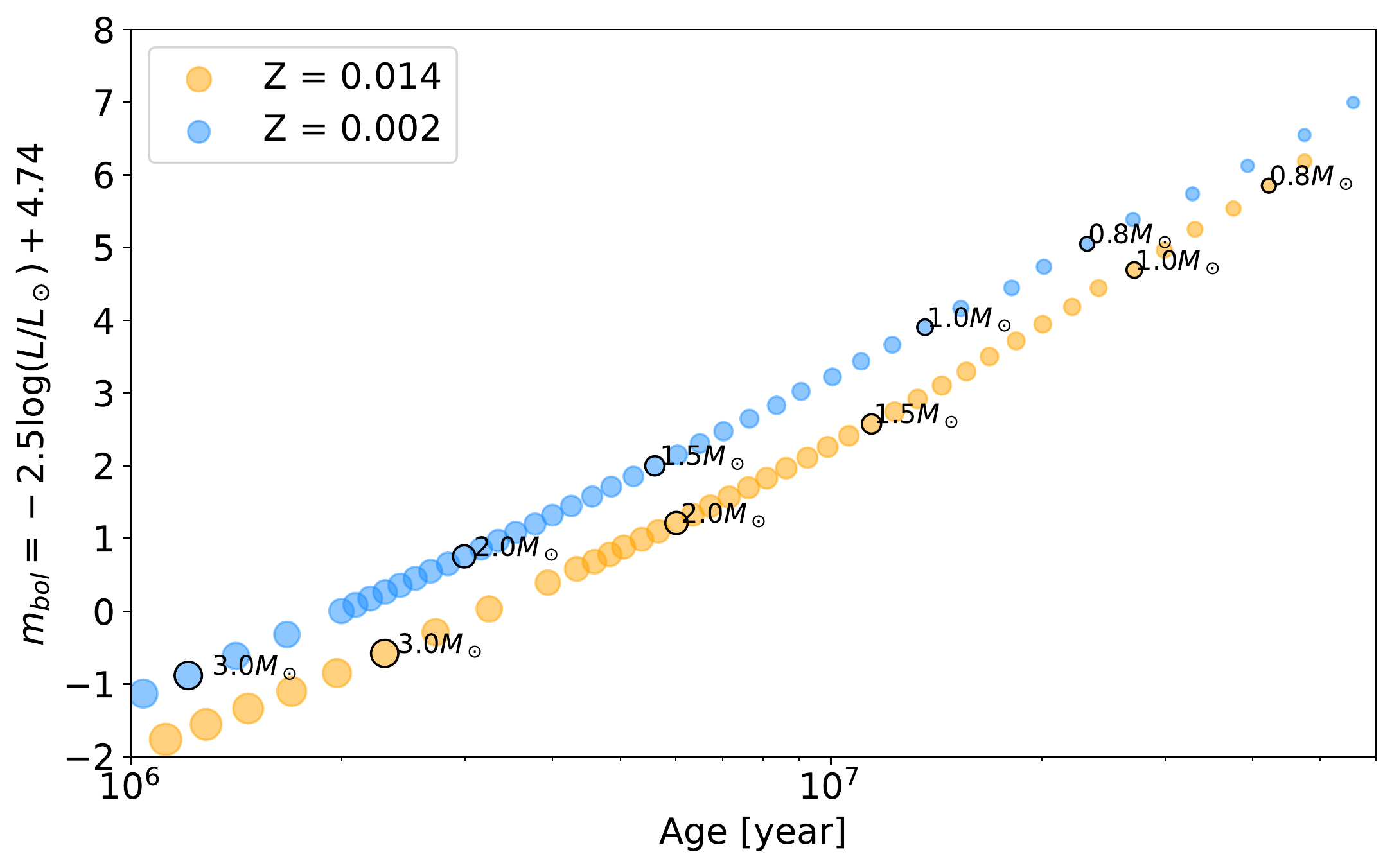}
                \caption{End PMS evolutionary critical points in bolometric magnitude as a function of age, in   Z = 0.002 (blue) and Z = 0.014 (orange). The size of the symbols is proportional to the corresponding initial mass of the model. The mass of some of the points is explicitly indicated in the plot with a black circle. This plot shows that at a given moment, in the lower metallicity population, the stars at the end of the PMS phase have a lower initial mass, and therefore a lower luminosity. This creates the shift we observe in the H-peak positions predicted by the lower metallicity models.}
                \label{fig:hpeakmbol-age-multiz}
        \end{figure}

        Figure~\ref{fig:hpeakmbol-age-multiz} shows the critical points of End PMS in the bolometric magnitude as a function of age, for both cases of Z. Each point represents a model of a certain initial mass at the phase of End PMS; the symbol size is proportional to the model's initial mass. For example, for the two models of 3~$M_\odot$ initial mass, the one with Z = 0.002 reaches the end of the PMS $\sim$ 1 Myr earlier than the one with Z = 0.014; both magnitudes are close to -1. For a lower initial mass, the time and magnitude discrepancy becomes larger. As a result, when we observe the H-peak in the luminosity function at a certain age of lower metallicity, its position is shifted towards the faint end because the population evolves faster; thus, the H-peak--age relation is shifted back in time. The faster PMS evolutionary speed is a consequence of a higher contraction rate in response to the lower metallicity, generating enough pressure to maintain the hydrostatic equilibrium. 
        
        \section{The color-absolute magnitude diagrams of the young populations}
\begin{figure*}
                \centering
                \includegraphics[width=0.8\linewidth]{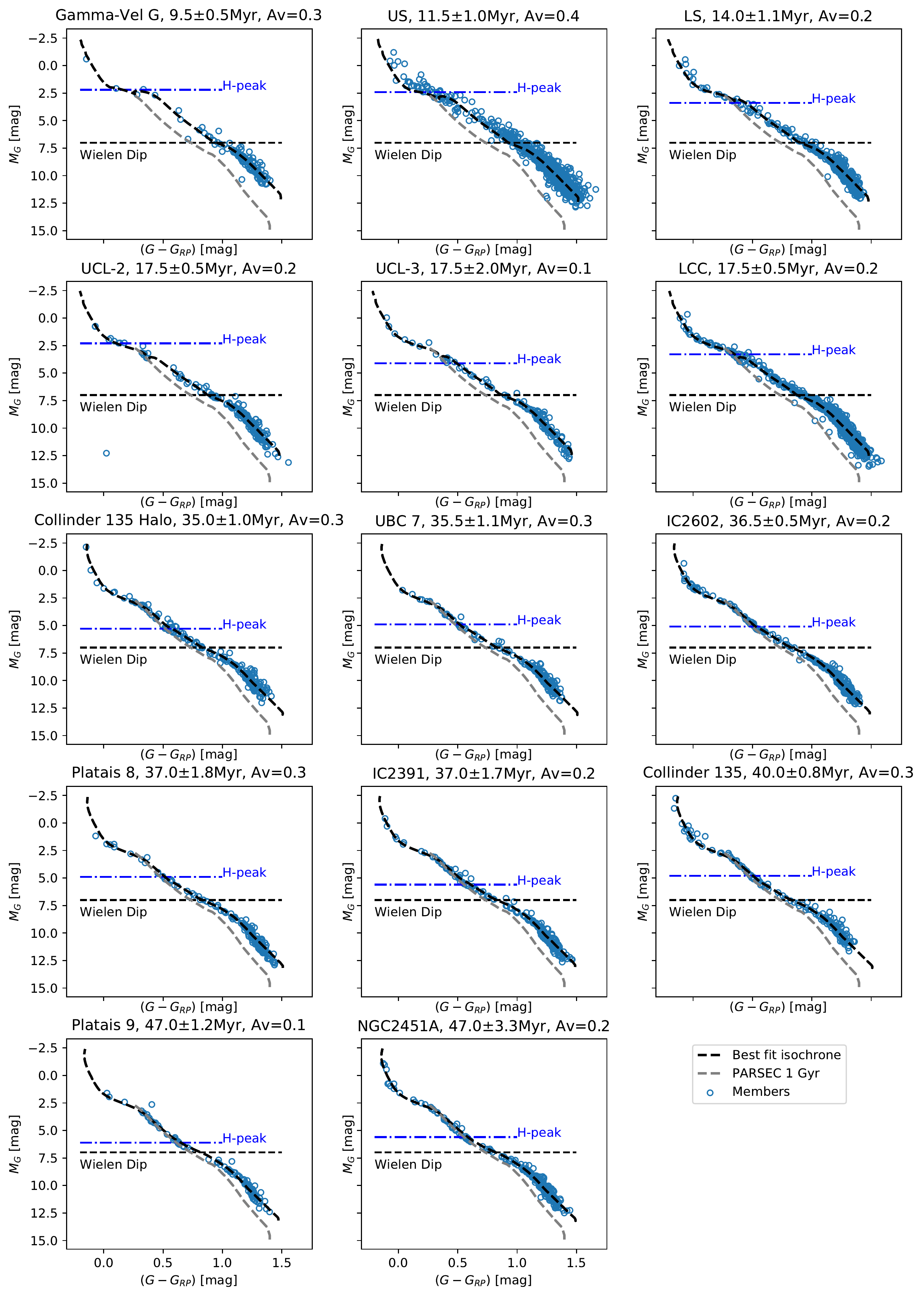}
                \caption{Color-absolute magnitude diagrams of the young populations with isochrones. The blue circles show the observed members. The error bars are smaller than the symbol size, and  the error bars are not plotted. The black dashed lines denote the best fit isochrone of the population. The gray dashed lines are the MS part of the 1 Gyr isochrone, which is used to approximate the position of the ZAMS on the intermediate- to low-mass end.
                The blue horizontal dot-dashed line indicates the observed mean location of the H-peak, the black dashed horizontal line the location of the Wielen dip at $M_G = 7$ mag.}
                \label{fig:cmdclusters}
        \end{figure*}

        Figure~\ref{fig:cmdclusters} shows the location of the H-peak and the Wielen dip in the color-absolute magnitude diagram for all investigated young clusters except for NGC 6231.
        
\end{document}